\documentclass[letter,apj]{emulateapj-rtx4}
\pdfoutput=1
\usepackage{graphicx}
\usepackage{enumerate}
\usepackage{amssymb, amsmath}
\usepackage{mathtools}
\usepackage{natbib}
\usepackage{color}
\usepackage{ulem}
\usepackage{soul}
\newcommand{\ias}{1}
\newcommand{\goddard}{2}
\newcommand{\princeton}{3}
\newcommand{\physmath}{4}
\newcommand{\tokyo}{5}
\newcommand{\naoj}{6}
\newcommand{\tokyotech}{7}
\newcommand{\mpia}{8}
\newcommand{\oklahoma}{9}
\newcommand{\fizeau}{10}
\newcommand{\edinburgh}{11}
\newcommand{\charleston}{12}
\newcommand{\toronto}{13}
\newcommand{\subaru}{14}
\newcommand{\sternwarte}{15}
\newcommand{\ifahawaii}{16}
\newcommand{\nagoya}{17}
\newcommand{\belfast}{18}
\newcommand{\sokendai}{19}
\newcommand{\hiroshima}{20}
\newcommand{\stsci}{21}
\newcommand{\jpl}{22}
\newcommand{\sinica}{23}
\newcommand{\ethz}{24}
\newcommand{\sapporo}{25}
\newcommand{\sendai}{26}

\begin{document}

\title{A Statistical Analysis of SEEDS and Other High-Contrast Exoplanet Surveys: Massive Planets or Low-Mass Brown Dwarfs?}
\author{Timothy D.~Brandt\altaffilmark{\ias}, 
Michael W.~McElwain\altaffilmark{\goddard}, 
Edwin L.~Turner\altaffilmark{\princeton,\physmath},
Kyle Mede\altaffilmark{\tokyo},
David S. Spiegel\altaffilmark{\ias},
Masayuki Kuzuhara\altaffilmark{\naoj,\tokyotech,\tokyo},
Joshua E.~Schlieder\altaffilmark{\mpia},
John~P. Wisniewski\altaffilmark{\oklahoma}, 
   L. Abe\altaffilmark{\fizeau},   
   B. Biller\altaffilmark{\edinburgh},
   W. Brandner\altaffilmark{\mpia},   
   J. Carson\altaffilmark{\charleston},    
   T. Currie\altaffilmark{\toronto},
   S. Egner\altaffilmark{\subaru},   
   M. Feldt\altaffilmark{\mpia},  
   T. Golota\altaffilmark{\subaru},   
   M. Goto\altaffilmark{\sternwarte},  
   C.~A. Grady\altaffilmark{\goddard}, 
   O. Guyon\altaffilmark{\subaru},  
   J. Hashimoto\altaffilmark{\naoj}, 
   Y. Hayano\altaffilmark{\subaru},  
   M. Hayashi\altaffilmark{\naoj},  
   S. Hayashi\altaffilmark{\subaru}, 
   T. Henning\altaffilmark{\mpia},   
   K.~W. Hodapp\altaffilmark{\ifahawaii},
   S. Inutsuka\altaffilmark{\nagoya},
   M. Ishii\altaffilmark{\subaru}, 
   M. Iye\altaffilmark{\naoj},    
   M. Janson\altaffilmark{\belfast},   
   R. Kandori\altaffilmark{\naoj},  
   G.~R. Knapp\altaffilmark{\princeton},   
   T. Kudo\altaffilmark{\naoj},   
   N. Kusakabe\altaffilmark{\naoj},   
   J. Kwon\altaffilmark{\naoj,\sokendai},
   T. Matsuo\altaffilmark{\naoj},   
   S. Miyama\altaffilmark{\hiroshima},   
   J.-I. Morino\altaffilmark{\naoj},  
   A. Moro-Mart\'in\altaffilmark{\stsci},  
   T. Nishimura\altaffilmark{\subaru},
   T.-S. Pyo\altaffilmark{\subaru}, 
   E. Serabyn\altaffilmark{\jpl},   
   H. Suto\altaffilmark{\naoj},    
   R. Suzuki\altaffilmark{\naoj},   
   M. Takami\altaffilmark{\sinica},  
   N. Takato\altaffilmark{\subaru}, 
   H. Terada\altaffilmark{\subaru},  
   C. Thalmann\altaffilmark{\ethz},  
   D. Tomono\altaffilmark{\subaru},  
   M. Watanabe\altaffilmark{\sapporo},  
   T. Yamada\altaffilmark{\sendai},   
   H. Takami\altaffilmark{\subaru},  
   T. Usuda\altaffilmark{\subaru}, 
   M. Tamura\altaffilmark{\tokyo,\naoj}  
}

\altaffiltext{*}{Based on data collected at Subaru Telescope, which
   is operated by the National Astronomical Observatory of Japan}
\altaffiltext{\ias}{Institute for Advanced Study, Princeton, NJ, USA}
\altaffiltext{\goddard}{Exoplanets and Stellar Astrophysics Laboratory, Goddard Space Flight Center, Greenbelt, MD, USA}
\altaffiltext{\princeton}{Department of Astrophysical Sciences, Princeton University, Princeton, NJ, USA}
\altaffiltext{\physmath}{Kavli Institute for the Physics and Mathematics of the Universe (WPI), Todai Institutes for Advanced Study, University of Tokyo, Tokyo, Japan}
\altaffiltext{\tokyo}{University of Tokyo, Tokyo, Japan}
\altaffiltext{\naoj}{National Astronomical Observatory of Japan, Tokyo, Japan}
\altaffiltext{\tokyotech}{Tokyo Institute of Technology, Tokyo, Japan}
\altaffiltext{\mpia}{Max Planck Institute for Astronomy, Heidelberg, Germany}
\altaffiltext{\oklahoma}{HL Dodge Department of Physics and Astronomy, University of Oklahoma, Norman, OK, USA}
\altaffiltext{\fizeau}{Laboratoire Hippolyte Fizeau, Nice, France}
\altaffiltext{\edinburgh}{University of Edinburgh, Edinburgh, Scotland, UK}
\altaffiltext{\charleston}{College of Charleston, Charleston, SC, USA}
\altaffiltext{\toronto}{Department of Astronomy and Astrophysics, University of Toronto, Toronto, ON, Canada}
\altaffiltext{\subaru}{Subaru Telescope, Hilo, Hawai`i, USA}
\altaffiltext{\sternwarte}{Universit\"ats-Sternwarte M\"unchen, Munich, Germany}
\altaffiltext{\ifahawaii}{Institute for Astronomy, University of Hawai`i, Hilo, Hawai`i, USA}
\altaffiltext{\nagoya}{Department of Physics, Nagoya University, Nagoya, Japan}
\altaffiltext{\belfast}{Queens University Belfast, Belfast, Northern Ireland, UKt}
\altaffiltext{\sokendai}{Department of Astronomical Science, Graduate University for Advanced Studies, Tokyo, Japan}
\altaffiltext{\hiroshima}{Hiroshima University, Higashi-Hiroshima, Japan}
\altaffiltext{\stsci}{Space Telescope Science Institute, Baltimore, MD, USA}
\altaffiltext{\jpl}{Jet Propulsion Laboratory, California Institute of Technology, Pasadena, CA, USA}
\altaffiltext{\sinica}{Institute of Astronomy and Astrophysics, Academia Sinica, Taipei, Taiwan}
\altaffiltext{\ethz}{Institute for Astronomy, ETH Z\"{u}rich, Z\"{u}rich, Switzerland}
\altaffiltext{\sapporo}{Department of Cosmosciences, Hokkaido University, Sapporo, Japan}
\altaffiltext{\sendai}{Astronomical Institute, Tohoku University, Sendai, Japan}

\begin{abstract}
We conduct a statistical analysis of a combined sample of direct imaging data, totalling nearly 250 stars.  The stars cover a wide range of ages and spectral types, and include five detections ($\kappa$ And b, two $\sim$60 M$_{\rm J}$ brown dwarf companions in the Pleiades, PZ Tel B, and CD$-$35 2722B).  For some analyses we add a currently unpublished set of SEEDS observations, including the detections GJ 504b and GJ 758B.  We conduct a uniform, Bayesian analysis of all stellar ages using both membership in a kinematic moving group and activity/rotation age indicators.  We then present a new statistical method for computing the likelihood of a substellar distribution function.  By performing most of the integrals analytically, we achieve an enormous speedup over brute-force Monte Carlo.  We use this method to place upper limits on the maximum semimajor axis of the distribution function derived from radial-velocity planets, finding model-dependent values of $\sim$30--100 AU.  Finally, we model the entire substellar sample, from massive brown dwarfs to a theoretically motivated cutoff at $\sim$5 M$_{\rm Jup}$, with a single power law distribution.  We find that $p(M, a) \propto M^{-0.65\pm0.60} a^{-0.85\pm0.39}$ (1$\sigma$ errors) provides an adequate fit to our data, with 1.0--3.1\% (68\% confidence) of stars hosting 5--70 $M_{\rm Jup}$ companions between 10 and 100 AU.  This suggests that many of the directly imaged exoplanets known, including most (if not all) of the low-mass companions in our sample, formed by fragmentation in a cloud or disk, and represent the low-mass tail of the brown dwarfs.
\end{abstract}

\section{Introduction} \label{sec:intro}

Since the first exoplanet around a main-sequence star was discovered in 1995 \citep{Mayor+Queloz_1995}, large radial velocity \citep[e.g.][]{Cumming+Butler+Marcy+etal_2008, Bonfils+Delfosse+Udry+etal_2013} and transit surveys \citep{Bakos+Noyes+Kovacs+etal_2004, Pollacco+Skillen+CollierCameron+etal_2006, Batalha+Rowe+Bryson+etal_2013} have found many hundreds of worlds.  Previous models of planet formation, extending back decades \citep[e.g.][]{Kuiper_1951, Hayashi_1981}, were based heavily on the Solar system.  New discoveries have enabled a much fuller characterization of exoplanets within $\sim$3 AU of their host stars, around both main sequence \citep{Cumming+Butler+Marcy+etal_2008} and evolved \citep{Johnson+Fischer+Marcy+etal_2007} systems.  These distributions hold clues to the formation and subsequent dynamical evolution of planetary systems.

While constraints on the mass function of planets have only recently been determined, the initial mass function (IMF) of stars has been studied for many decades \citep{Salpeter_1955}, and is now well-constrained.  The stellar IMF has also recently been extended to brown dwarfs \citep{Kroupa_2001, Reid+Gizis+Hawley_2002, Chabrier_2003}, objects below the minimum mass ($\sim$80 M$_{\rm J}$) necessary to sustain hydrogen fusion, but above the minimum mass for deuterium burning ($\sim$13 M$_{\rm J}$).  Large samples of substellar objects are difficult to assemble, both because brown dwarfs have a limited supply of internal energy, and because the IMF turns over near the hydrogen-burning boundary.  Brown dwarfs are also uncommon as close companions to main-sequence stars, a phenomenon known as the ``brown dwarf desert'' \citep{Marcy+Butler_2000, Grether+Lineweaver_2006}.  The companion mass function (CMF) rises from this ``desert'' both towards higher, stellar masses, and towards lower, planetary masses. 

The companion mass function, well-established at small separations from radial velocity surveys \citep{Cumming+Butler+Marcy+etal_2008}, is much less clear at tens of AU.  The conditions in a protoplanetary disk are very different far from the host star and may not support the formation mechanism responsible for the radial-velocity (RV) distribution \citep{Dodson-Robinson+Veras+Ford+etal_2009}, though such a conclusion is far from certain \citep{Lambrechts+Johansen_2012, Kenyon+Bromley_2009}.  Companions near or below the deuterium-burning limit might form like stars in gravitational collapse or fragmentation \citep{Boss_1997, Vorobyov_2013}, by core-accretion in-situ \citep{Pollack+Hubickyj+Bodenheimer_1996, Alibert+Mordasini+Benz+etal_2005}, or they might form in a location more conducive to planet formation and subsequently migrate or be scattered to their observed orbits.  The distribution function of such objects could provide important clues to their formation mechanism, and connect them to either more massive brown dwarfs or to the planet populations observed at smaller separations.  As a result, these massive exoplanets are being heavily targeted using large telescopes with adaptive optics.

The sensitivity to exoplanets with direct imaging has been rapidly improving, and recent discoveries have begun to fill the parameter space of substellar objects at separations of tens of AU.  Companions near or below the deuterium burning limit have been discovered around the M-dwarfs 2MASS J01225093-2439505 \citep{Bowler+Liu+Shkolnik+etal_2013}, 2MASS J01033563-5515561AB \citep{Delorme+Gagne+Girard+etal_2013}, and ROXs 42B \citep{Currie+Daemgen+Debes+etal_2014}, the G-dwarf GJ 504 \citep{Kuzuhara+Tamura+Kudo+etal_2013}, the A stars HR 8799 \citep{Marois+Macintosh+Barman+etal_2008, Marois+Zuckerman+Konopacky+etal_2010}, $\beta$ Pictoris \citep{Lagrange+Gratadour+Chauvin+etal_2009}, and HD 95086 \citep{Rameau+Chauvin+Lagrange+etal_2013}, and the late B star $\kappa$ And \citep{Carson+Thalmann+Janson+etal_2013}.  These detections were made possible by recent technological advances in adaptive optics and the use of differential imaging techniques.  In addition, recent work to identify nearby members of young moving groups \citep[MGs, e.g.][]{Torres+Quast+Melo+etal_2008, Zuckerman+Rhee+Song+etal_2011, Schlieder+Lepine+Simon_2012a, Shkolnik+Anglada+Liu+etal_2012, Malo+Doyon+Lafreniere+etal_2013, Moor+Szabo+Kiss+etal_2013, Rodriguez+Zuckerman+Kastner+etal_2013, Gagne+Lafreniere+Doyon+etal_2014} and stars that harbor debris disks \citep{Rieke+Su+Stansberry+etal_2005, Chen+Sargent+Bohac+etal_2006, Plavchan+Werner+Chen+etal_2009, Moor+Pascucci+Kospal+etal_2011, Eiroa+Marshall+Mora+etal_2011} has provided excellent targets for direct imaging searches.  The stellar age is particularly important, because substellar objects are unable to sustain hydrogen fusion in their cores, and quickly fade beneath the sensitivity limits of the best observing facilities on the ground and in space.  

Over the last decade, numerous direct imaging surveys around nearby young stars have begun to constrain the distribution and frequency of substellar companions.  These have mostly used non-detections to place upper limits on the planet fraction within a range of masses and semimajor axes, or upper limits beyond which the distribution function measured by radial velocity surveys cannot extend.  \cite{Lafreniere+Doyon+Marois+etal_2007} used the Gemini Deep Planet Survey (GDPS) to place upper limits of $\sim$10\% on the fraction of stars with 0.5--13 M$_{\rm J}$ companions in the range from 50 to 250 AU, assuming an RV-like mass distribution.  \cite{Nielsen+Close_2010} used a sample of 118 targets, dominated by the GDPS, to find that the RV distribution of \cite{Cumming+Butler+Marcy+etal_2008} cannot be extrapolated past $a_{\rm max}$ from $\sim$65--200 AU depending on the substellar cooling model and on the correlation between planet frequency and stellar mass.  \cite{Chauvin+Lagrange+Bonavita+etal_2010} imaged 88 stars, around which they detected three substellar companions, including an $\sim$8-M$_{\rm J}$ companion to the brown dwarf 2M1207 and a $\sim$13 M$_{\rm J}$ companion to AB Pic.  More recently, \cite{Vigan+Patience+Marois+etal_2012} placed a lower limit to the planet ($<$13-M$_{\rm J}$) frequency around A stars of 6\% based on the International Deep Planet Survey and the detections around HR 8799 and $\beta$ Pictoris.  \cite{Biller+Liu+Wahhaj+etal_2013} placed a similar model-dependent upper limit of 6\%--18\% for companions from 1--20 M$_{\rm J}$ between 10 and 150 AU around later-type stars.  \cite{Chauvin+Vigan+Bonnefoy+etal_2014} observed 86 stars without detecting substellar companions, placing an upper limit of 10\% on 5--10 $M_{\rm Jup}$ objects 50--500 AU from young solar-type stars.  However, the distribution function for these companions remains uncertain, and statistical analyses often artificially truncate it at or near the deuterium burning threshold.  

In this work, we provide a new framework for determining the distribution function of substellar companions to stars, and apply this framework to the published sub-sample of the Subaru SEEDS survey, combined with the publicly available GDPS \citep{Lafreniere+Doyon+Marois+etal_2007} and NICI MG sample \citep{Biller+Liu+Wahhaj+etal_2013}.  In Section \ref{sec:distributions}, we discuss what is currently known about the stellar and substellar mass distributions.  In Section \ref{sec:data}, we present our combined data set, and in Section \ref{sec:bayes_ages}, we summarize our method for deriving their age probability distributions; in Section \ref{sec:cooling_models}, we discuss our choice of substellar cooling models.  We present our statistical framework and method for determining constraints on the substellar distribution function in Section \ref{sec:framework}, with additional details, derivations, and fitting functions in the Appendix.  Section \ref{sec:results} presents and discusses our results, including our limits on an extrapolated RV-like distribution function, and the ability of a single distribution to include most or all wide-separation substellar objects.  We conclude with Section \ref{sec:conclusions}.

\section{A Tail of Two Distributions} \label{sec:distributions}

In spite of their rarity, brown dwarf and very massive ($\gtrsim$10 $M_{\rm Jup}$) giant planet companions are now being discovered at wide separations around nearby stars, both by dedicated high-contrast surveys and by mid-infrared space missions such as the {\it Wide-field Infrared Survey Explorer} ({\it WISE}).  In contrast, less-massive exoplanets at wide separations remain extremely scarce:  only two known companions to stellar primaries are likely to have masses less than 5 $M_{\rm Jup}$: GJ 504b \citep{Kuzuhara+Tamura+Kudo+etal_2013}, and HD 95086b \citep{Rameau+Chauvin+Lagrange+etal_2013}.  LkCa 15b \citep{Kraus+Ireland_2012} and, especially, Fomalhaut b \citep{Kalas+Graham+Chiang+etal_2008, Janson+Carson+Lafreniere+etal_2012, Currie+Debes+Rodigas+etal_2012} may also fall into this category, though their natures remain unclear.  The paucity of known $\sim$3--5 M$_{\rm J}$ companions might indicate a lack of planets rather than simply our inability to detect them.  High-contrast surveys have now observed $\sim$1000 nearby stars, often with sensitivities to objects significantly less massive than those detected around $\beta$ Pictoris and HR 8799 at moderate to wide separations ($\gtrsim 1''$).  

The mass distribution function of short-period companions, as determined by radial velocity surveys, is now well-determined.  It decreases sharply from $\sim$1 M$_{\rm J}$ to $\sim$20 M$_{\rm J}$, then increases towards stellar masses \citep{Marcy+Butler_2000, Grether+Lineweaver_2006}.  \cite{Cumming+Butler+Marcy+etal_2008}, using eight years of radial velocity data and assuming $dN/dM \propto M^{\beta}$, find $\beta_{\rm pl} = -1.31 \pm 0.2$ for $M_{\rm pl} > 0.3$ M$_{\rm J}$ and periods $<$2000 days.  The relative lack of companions in the mass range from $\sim$10 M$_{\rm J}$ to $\sim$40 M$_{\rm J}$ is known as the brown dwarf desert.  This ``desert,'' however, is significantly less pronounced at wider separations \citep{Gizis+Kirkpatrick+Burgasser+etal_2001, Metchev+Hillenbrand_2009}.  

Recent discoveries have illuminated the brown dwarf mass function in the field, favoring a power law index $\beta_{\rm BD} \sim 0$ for a mass distribution of the form $dN/dM \propto M^{\beta}$.  The mass distribution is certainly not an extension of that of low-mass stars \citep{Kroupa_2001, Reid+Gizis+Hawley_2002, Chabrier_2003}, and the best-fitting piecewise power law may be discontinuous \citep{Thies+Kroupa_2007}.  \cite{Thies+Kroupa_2008} find a best fit $\beta_{\rm BD} \sim -0.3$ using young clusters, while \cite{Kirkpatrick+Gelino+Cushing+etal_2012} find $0 \lesssim \beta_{\rm BD} \lesssim 0.5$ in the very nearby field, and \cite{Sumi+Kamiya+Bennett+etal_2011} report $\beta_{\rm BD} = -0.49^{+0.27}_{-0.24}$ based on gravitational microlensing events.  \cite{Sumi+Kamiya+Bennett+etal_2011} hypothesized a distribution of low-mass ``rogue planets'' in the field to explain their shortest duration microlensing events, and found $\beta = -1.3^{+0.4}_{-0.3}$ for such a population, consistent with that found for radial velocity companions.  A value $\beta_{\rm BD} = 0$ corresponds to equal numbers of objects at all brown dwarf masses, with most of the matter residing in massive objects.  

The brown dwarf mass function in binaries is more controversial.  \cite{Metchev+Hillenbrand_2009}, in a high-contrast imaging survey of young stars, found their data to be consistent with a single power law companion mass function with $\beta \approx -0.4$ extending all the way from 1 M$_{\odot}$ down to the deuterium-burning limit of $\sim$13 M$_{\rm J}$.  This result is strongly inconsistent with the distribution of companion masses seen at small separations.  In contrast, \cite{Zuckerman+Song_2009} derived the brown dwarf companion mass function from a survey of objects reported in the literature, and found a much steeper $\beta \sim -1.2$.  However, hydrodynamical simulations disfavor such a strong peak at low masses, with \cite{Stamatellos+Whitworth_2009} finding a relatively flat distribution of cloud fragment masses over the brown dwarf regime.  

As with the mass distribution, the semimajor axis distributions for stellar binaries and for close ($\lesssim$3 AU) exoplanets are both well-determined.  \cite{Cumming+Butler+Marcy+etal_2008} find $\alpha = -0.61 \pm 0.15$ for $dN/da \propto a^\alpha$ in the range from 0.03--3 AU, i.e., a distribution with planet incidence decreasing with separation, but still divergent if integrated to infinity.  The semimajor axis distribution of companions to G-dwarfs is approximately log-normal, with a peak in the distribution at $\sim$30--50 AU for a star of 1 M$_\odot$ \citep{Duquennoy+Mayor_1991, Raghavan+McAlister+Henry+etal_2010}.  In the range $5 < a < 500$ AU, the space most relevant to high-contrast imaging, the observed distribution is very nearly flat in $\log a$ (varying by $\sim$30\%).

The mass (and separation) ranges over which the planet and brown dwarf distributions apply are controversial.  The IAU draft definition of a planet is simply any object which is both bound to a star and below the minimum mass for deuterium burning, currently calculated to be $\sim$13 M$_{\rm J}$ \citep{Spiegel+Burrows+Milsom_2011}.  However, this criterion holds no other physical significance and is independent of the object's formation mechanism.  Brown dwarfs are thought to form like stars, and, though the minimum fragment mass is not known precisely, it is likely to fall at around 5 M$_{\rm J}$ \citep{Low+Lynden-Bell_1976, Bate_2003, Bate_2009}.  Some studies \citep{Bate_2003, Umbreit+Burkert+Henning+etal_2005} suggest that brown dwarfs form when they are ejected from dynamically unstable systems, making it extremely difficult to form substellar companions to solar-type stars by cloud fragmentation.  However, in the outer regions of a disk rather than in a molecular cloud, fragmentation may be more likely to produce substellar objects \citep{Boss_1997}, though the lower mass cutoff is still likely to be several M$_{\rm J}$ \citep{Rafikov_2005, Whitworth+Stamatellos_2006}.  

Substellar objects near or below the deuterium burning limit of $\sim$13 M$_{\rm J}$ have been discovered at separations from $\sim$40 AU to $\sim$1000 AU around stars ranging in spectral type from late M to late B \citep[e.g.][]{Chauvin+Lagrange+Zuckerman+etal_2005, Marois+Macintosh+Barman+etal_2008, Lagrange+Gratadour+Chauvin+etal_2009, Bowler+Liu+Shkolnik+etal_2013, Carson+Thalmann+Janson+etal_2013, Currie+Daemgen+Debes+etal_2014}.  Two recent discoveries, GJ~504b \citep{Kuzuhara+Tamura+Kudo+etal_2013} and HD~95086b \citep{Rameau+Chauvin+Lagrange+etal_2013}, plus HR 8799b \citep{Sudol+Haghighipour_2012}, push the mass range of such companions down to $\sim$5 M$_{\rm J}$.  If these objects could form by core-accretion \citep{Pollack+Hubickyj+Bodenheimer_1996, Alibert+Mordasini+Benz+etal_2005, Lambrechts+Johansen_2012}, believed to be responsible for gas giants at smaller separations, they might be an extension of the radial velocity distribution (nearly all planets of which, presumably, formed via core accretion).  If, however, they form by fragmentation in a cloud or disk \citep[e.g.][]{Boss_1997, Vorobyov_2013}, they should be considered together with low-mass brown dwarfs, and perhaps with the entire substellar distribution up to a mass of $\sim$0.1 M$_\odot$, where the initial mass function turns over and starts rising toward higher mass.  It is also possible that more than one of these distributions overlap in parameter space, and that the sample of direct-imaging discoveries is a heterogeneous population with multiple modes of formation \citep{Boley_2009}.

Because many discoveries straddle the deuterium burning threshold (depending on the assumed age of the system and on the substellar cooling model), and because there is no theoretical motivation to separate objects at 13 M$_{\rm J}$, we consider giant planets and low-mass brown dwarfs together.  It is certainly possible that the most massive brown dwarfs ($\gtrsim$50 M$_{\rm J}$) are drawn from a different distribution, but we see no reason to impose such a division {\it \`a priori}, particularly not at 13 M$_{\rm J}$.  The form of the distribution function and its limits in mass and semimajor axis should be powerful clues to the formation mechanism at work.  We note, however, that we are still limited by our reliance on uncertain models of substellar cooling and by our parametrization of the distribution function with limited theoretical motivation.

\section{High Contrast Imaging Data Set} \label{sec:data}

We merge five samples to create our high-contrast data set: three published subgroups of the Strategic Exploration of Exoplanets and Disks with Subaru project \citep[SEEDS, ][]{Tamura_2009}, the Gemini Deep Planet Survey \citep[GDPS, ][]{Lafreniere+Doyon+Marois+etal_2007}, and the MG targets from the NICI survey \citep{Biller+Liu+Wahhaj+etal_2013}.  We discuss each of these in turn.  In total, our merged survey contains 248 unique stars with spectral types ranging from late B to mid M, at distances from $\sim$5 pc to $\sim$130 pc, with sensitivities down to $\sim$1 M$_{\rm J}$ around the nearest, youngest targets.  We reduce most of the data, including that of GDPS targets but excepting the NICI stars, using ACORNS-ADI \citep{Brandt+McElwain+Turner+etal_2013}; for GDPS stars, our results are very similar to those published in \cite{Lafreniere+Doyon+Marois+etal_2007}.  For the NICI data, we use the published reductions as described by \cite{Wahhaj+Liu+Biller+etal_2013}.

After reviewing each of the five samples, we apply a uniform, Bayesian analysis to compute age probability distributions.  This analysis combines proposed MG membership, rotation and activity indicators, all of which are listed in Table \ref{tab:age_indicators}.  We provide a summary in Section \ref{sec:bayes_ages}, with the full procedure described in \cite{Brandt+Kuzuhara+McElwain+etal_2014}.  The secondary age indicators of the SEEDS MG sample are listed in \cite{Brandt+Kuzuhara+McElwain+etal_2014}, while the indicators for the other samples are listed in Table \ref{tab:age_indicators}.  Our age analysis also requires us to estimate the probability of MG membership for each target.  Most of our targets are either consensus members of a MG (listed as 95\%$+$ probability in Table \ref{tab:age_indicators}) or consensus nonmembers, though a few are more uncertain.  We briefly discuss each such case, including our adopted MG classification and membership probability, in Section \ref{subsec:uncertain_mg}.

In addition to the 248 stars in this sample, we consider two additional stars around which HiCIAO has detected substellar companions: GJ 758B \citep{Thalmann+Carson+Janson+etal_2009, Janson+Carson+Thalmann+etal_2011}, a $\sim$30 M$_{\rm J}$ brown dwarf around an old G star first imaged during HiCIAO commissioning, and GJ 504b \citep{Kuzuhara+Tamura+Kudo+etal_2013, Janson+Brandt+Kuzuhara+etal_2013}, a $\sim$3--8 M$_{\rm J}$ companion to an active field G star discovered during the full SEEDS survey.  By using these discoveries in part of our statistical analysis, we assume that the stars and contrast curves of $\sim$200 as-yet-unpublished SEEDS targets are similar to those of the merged sample we discuss in this section.  This is a fairly good approximation; in fact, the unpublished SEEDS targets are primarily a combination of relatively distant ($\sim$100 pc), very young ($\sim$5--10 Myr) pre-main-sequence stars, and very nearby stars ($\sim$5--50 pc) with a wide range of ages (tens of Myr to several Gyr) and spectral types (A through M).

\subsection{SEEDS} \label{subsec:seeds_data}

SEEDS \citep{Tamura_2009} is now mostly complete, having observed nearly 400 nearby stars with the high-contrast camera HiCIAO \citep{Suzuki+Kudo+Hashimoto+etal_2010} behind the 188-actuator adaptive optics system AO188 \citep{Hayano+Takami+Guyon+etal_2008}.  For this study, we use the previously published data sets comprised of debris disk hosts \citep{Janson+Brandt+Moro-Martin+etal_2013}, proposed members of nearby MGs \citep{Brandt+Kuzuhara+McElwain+etal_2014}, and Pleiades members \citep{Yamamoto+Matsuo+Shibai+etal_2013}.  The Pleiades targets are young, with ages that are better-constrained than those of typical field stars.  

The SEEDS MG sample is described in detail in \cite{Brandt+Kuzuhara+McElwain+etal_2014}.  That work includes stellar age indicators and Bayesian age estimates for all targets, both those that are reliable members of coeval MGs and those that are not.  We adopt those indicators and ages throughout the rest of this work.  There is much less controversy over membership in the Pleiades, a rich, young \citep[$125 \pm 8$ Myr;][]{Stauffer+Schultz+Kirkpatrick_1998} stellar cluster at a distance of $\sim$130 pc.  There is, however, considerable controversy over the distance to the Pleiades, with {\it Hipparcos} parallaxes favoring a value of 120 pc \citep{vanLeeuwen_2009}, against isochrone fitting and other techniques giving a value about 10\% larger \citep{An+Terndrup+Pinsonnealt+etal, Munari+Dallaporta+Siviero+etal_2004}.  These differences are of marginal significance to our work.  We adopt a distance of $130 \pm 10$ pc for all imaged Pleiades targets, together with a slightly more conservative age of $125 \pm 10$ Myr.  

The MG sample includes one substellar detection, $\kappa$ And b \citep{Carson+Thalmann+Janson+etal_2013}.  The object lies at a projected separation of 55 AU, and has a mass of $\sim$13 M$_{\rm J}$ assuming the primary, $\kappa$ And, to be a member of the $\sim$30-Myr Columba MG.  Recent papers have called its Columba membership into question due to the star's position on the color-magnitude diagram, which hints at an age of several hundred Myr and a mass of $\sim$50 M$_{\rm J}$ for the companion \citep{Hinkley+Pueyo+Faherty+etal_2013, Bonnefoy+Currie+Marleau+etal_2014}.  The Pleiades targets include two substellar detections: {\sc H\,ii} 1348B \citep{Geissler+Metchev+Pham+etal_2012} and HD 23514B \citep{Rodriguez+Marois+Zuckerman+etal_2012}.  {\sc H\,ii} 1348B and HD 23514B lie at projected separations of 140 and 310 AU respectively, and each has a mass of $\sim$60 M$_{\rm J}$.  Neither discovery was published when the SEEDS observations were made. 

The debris disk sample has its stellar properties listed in Table 4 of \cite{Janson+Brandt+Moro-Martin+etal_2013}.  For this work, we revisit the ages using the Bayesian techniques presented in \cite{Brandt+Kuzuhara+McElwain+etal_2014} and summarized in Section \ref{sec:bayes_ages}.  

\subsection{GDPS} \label{subsec:gdps_data}

GDPS \citep{Lafreniere+Doyon+Marois+etal_2007} used the NIRI instrument on Gemini-North to take high-contrast near-infrared images of 85 nearby stars.  The stars ranged from F to M dwarfs, and were selected to be young, as inferred from their membership in kinematic MGs or from fits to color-magnitude isochrones.  As for our other data sets, we generally do not consider isochrone ages.  These are typically unreliable for main-sequence stars that have completed $\lesssim 1/3$ of their main-sequence lives \citep[][ and references therein]{Soderblom_2010}, precisely those that we are most interested in dating.  We do consider isochrones in some cases where a MG identification is in doubt \citep{Brandt+Kuzuhara+McElwain+etal_2014}.  GDPS did not detect any substellar companions to its targets, though the survey did find several new stellar binaries.

\subsection{NICI} \label{subsec:nici_data}

We also include the unique stars added by the southern sample observed by \cite{Biller+Liu+Wahhaj+etal_2013} using the NICI instrument on Gemini-South.  To enable a straightforward comparison to the data from the other surveys, we consider only those stars observed with angular differential imaging in the $H$-band, excluding those targets with spectral differential imaging (SDI).  \cite{Biller+Liu+Wahhaj+etal_2013} also report 95\% completeness levels rather than 5$\sigma$ (or 5.5 or 6$\sigma$) thresholds, where $\sigma$ is the standard deviation of the background at a given separation.  In order to convert these levels into a common framework with the values reported by other surveys, we assume the residuals to be Gaussian (a very good approximation for SEEDS and GDPS).  A 95\% completeness threshold, assuming a 5$\sigma$ minimum for follow-up, corresponds to a source bright enough to be detected if it were located on 95\% of the background fluctuations.  Given that a Gaussian distribution with zero mean has 95\% of its probability above $-1.64$$\sigma$, a 6.64$\sigma$ source would be detected at 5$\sigma$ or better 95\% of the time.
We therefore convert the 95\% limits to 5$\sigma$ sensitivity curves by adding
\begin{equation}
2.5\log_{10} \frac{6.64}{5} \approx 0.31~{\rm mag}~,
\end{equation}
and proceed to include the modified sensitivity curves in the remainder of our analysis.  The NICI sample includes two substellar detections: a $\sim$35 M$_{\rm J}$ brown dwarf around HIP 92680  \citep[= PZ Tel,][]{Biller+Liu+Wahhaj+etal_2010}, and a $\sim$30 M$_{\rm J}$ brown dwarf around TYC 7084-794-1 \citep[= CD$-$35 2722,][]{Wahhaj+Liu+Biller+etal_2011}.  HD 1160 was not in the NICI MG sample, so its brown dwarf companion \citep{Nielsen+Liu+Wahhaj+etal_2012} does not enter our analysis.

\subsection{Targets with Uncertain MG Membership} \label{subsec:uncertain_mg}

Most of the stars in our combined sample are either very high probability members of a given MG (95\%$+$ in Table \ref{tab:age_indicators}), or are unlikely to be members of any known MG.  We generally consider a star to be an uncontroversial MG member if it has a 99\% or higher membership probability as given by \cite{Malo+Doyon+Lafreniere+etal_2013}, or a 95\% or higher membership probability according to \cite{Gagne+Lafreniere+Doyon+etal_2014} (who updated \citeauthor{Malo+Doyon+Lafreniere+etal_2013} to include non-uniform priors), together with at least one additional youth indicator.  The Bayesian analyses of \cite{Malo+Doyon+Lafreniere+etal_2013} and \cite{Gagne+Lafreniere+Doyon+etal_2014} include position and proper motion and, where available, radial velocity and parallax (giving a position in six-dimensional phase space).  Not all of the high probability members have radial velocity and parallax measurements (e.g.~most TW Hydrae stars were too faint for {\it Hipparcos}), but a Bayesian analysis accounts for this.

This section summarizes the information for those stars which fall short of our criteria for near-certain MG membership, but are still possible or likely members of a MG.  All stars in the following summary have position, proper motion, parallax, and radial velocity measurements unless otherwise noted. 

\subsubsection{SEEDS Debris Disk Targets}

{\it HIP 7345}---\cite{Zuckerman+Song_2012} identify this early A star as a member of the Argus association, a classification definitively rejected by {\sc banyan-ii}\footnote{See http://www.astro.umontreal.ca/\textasciitilde malo/banyan.php, http://www.astro.umontreal.ca/\textasciitilde gagne/banyanII.php} at 99\% probability.  We decline to identify the star with any association.

{\it HIP 7576}---This mid-G star was classified in the Hercules-Lyra association, which we consider to be unreliable for dating \citep{Brandt+Kuzuhara+McElwain+etal_2014}.  It does show abundant youth indicators, and its kinematics and parallax also fit very nicely with the predictions for $\beta$ Pic membership; {\sc banyan} places it in $\beta$ Pic with over 99\% probability.  We provisionally adopt the $\beta$ Pic classification here, though the secondary age indicators are slightly more consistent with an age of several hundred Myr, rather than the $\sim$20 Myr of $\beta$ Pic.

{\it HIP 11360}---This early F star is listed as a bona-fide member with ambiguous group membership \citep{Malo+Doyon+Lafreniere+etal_2013}.  \cite{Torres+Quast+daSilva+etal_2006} suggest membership in $\beta$ Pic, while \cite{Malo+Doyon+Lafreniere+etal_2013} find a better fit with membership in Columba; the {\sc banyan ii} analysis (without a radial velocity) also places a significant ($\sim$15\%) probability of field membership.  We provisionally adopt the Columba classification, with its slightly older age, at 85\% probability.

{\it HIP 11847}---The {\sc banyan} web tool lists a 99\% probability of $\beta$ Pic membership for this early F star (with radial velocity from \citealt{Moor+Pascucci+Kospal+etal_2011}), though {\sc banyan ii} finds a probability of just $\sim$13\%.  \cite{Moor+Pascucci+Kospal+etal_2011} also note a nearby star with similar space motions and strong youth indicators, supporting $\beta$ Pic membership for both stars.  HIP 11847 hosts a bright debris disk \citep{Moor+Pascucci+Kospal+etal_2011}, and its isochrone-based age likelihood \citep[see ][ for details]{Brandt+Kuzuhara+McElwain+etal_2014} is consistent with an age from $\sim$20--200 Myr.  For this work, we adopt a 50\% probability of $\beta$ Pic membership. 

{\it HIP 13402}---{\sc banyan} gives a $\sim$70\% membership probability in $\beta$ Pic for this early K star; {\sc banyan ii} gives 22\%.  It has abundant secondary age indicators, including vigorous X-ray and chromospheric activity, rapid rotation, and abundant lithium, all consistent with a very young age.  We adopt the lower probability here.

{\it HIP 57632}---\cite{Zuckerman+Rhee+Song+etal_2011} classified this early A star as a member of the Argus association, which {\sc banyan ii} confirms, with a probability of just over 90\%.  We accept Argus membership with the latter's significance.

{\it HIP 76829}---This F4 star lacks any youth indicators apart from its kinematics, and has not been proposed as a bona fide member of any kinematic MG apart from the Hercules-Lyra association \citep{Lopez-Santiago+Montes+Crespo-Chacon+etal_2006}, which we consider to be unreliable for dating \citep{Brandt+Kuzuhara+McElwain+etal_2014}.  The star is too massive to show strong magnetic activity.  Its position on the color-magnitude diagram appears to be consistent with either an age of $\sim$20 Myr, consistent with $\beta$ Pic, or several hundred Myr.  For this analysis, we adopt {\sc banyan ii}'s 84\% membership probably in $\beta$ Pic.  

{\it HIP 77542}---This B9.5 star is a pre-main-sequence object with a relatively gas-rich disk \citep{Merin+Montesinos+Eiroa+etal_2004}, marking it as a very young object.  We follow \cite{Merin+Montesinos+Eiroa+etal_2004} in adopting an age of $5 \pm 1$ Myr.

{\it HIP 79977}---This F2 star is a consensus member of the Upper Scorpuis OB association \citep{deZeeuw+Hoogerwerf+deBruijne+etal_1999}.  We adopt an age of $11 \pm 3$ Myr for the association, consistent with two recent estimates \citep{Pecaut+Mamajek+Bubar_2012, Song+Zuckerman+Bessell_2012}.  

\subsubsection{non-SEEDS GDPS Targets}

{\it HIP 9291}---{\sc Banyan ii} places this mid-M star in Carina with $\sim$75\% probability.  It does show vigorous X-ray activity and is rapidly rotating, with $v \sin i \sim 15$ km\,s$^{-1}$ \citep{Glebocki+Gnacinski_2005}.  Given these signs of youth, we provisionally accept the 75\% probability estimate of Columba membership.

{\it HIP 44458}---This G0 star has very strong secondary youth indicators, including strong lithium absorption and rapid rotation.  While {\sc banyan} gives a 70\% probability of membership in Columba, {\sc banyan ii} gives just 3\%.  We decline to identify the star with any MG, relying instead on its abundant secondary age indicators.  

{\it HIP 72339}---{\sc banyan} places this star in AB Dor with 96\% probability based on its kinematics; {\sc banyan ii} gives a 24\% probability of AB Dor membership.  However, none of its youth indicators are consistent with such a young age; its lithium and activity levels are more consistent with an age of $\sim$1 Gyr.  The isochrone-based likelihood shows two peaks, one around $\sim$50 Myr, and another around 1 Gyr.  We therefore provisionally reject AB Dor membership.  

{\it HIP 105038}---{\sc banyan} places this K3 star in either Columba or $\beta$ Pic with high probability based on its kinematics.  However, its youth indicators favor a much older age, and {\sc banyan ii} places it in the field with over 90\% probability.  Similarly to HIP 72339, we provisionally reject Columba membership.  

{\it HIP 113020}---This mid-M star, GJ 876, is known to host four planets discovered by radial velocity \citep{Delfosse+Forveille+Mayor+etal_1998, Marcy+Butler+Vogt+etal_1998, Marcy+Butler+Fischer+etal_2001, Rivera+Lissauer+Butler+etal_2005, Rivera+Laughlin+Butler+etal_2010}.  It has only a very weak upper limit on X-ray activity and a measurement of slow rotation \citep[$\sim$2 km\,s$^{-1}$, ][]{Glebocki+Gnacinski_2005} to corroborate {\sc banyan}'s 99.9\% classification in $\beta$ Pic (85\% in {\sc banyan ii}).  \cite{Marcy+Butler+Fischer+etal_2001} estimate HIP 113020's age to be $>$1 Gyr due to its lack of activity, which would be exceptional in a $\sim$20 Myr-old $\beta$ Pic member.  We decline to identify the star with any moving group.

{\it HIP 115147}---This K1 star shows very strong chromospheric and X-ray activity, rapid rotation \citep[$\sim$16 km\,s$^{-1}$, ][]{Glebocki+Gnacinski_2005}, and retains abundant lithium, all consistent with a young age.  We therefore accept {\sc banyan}'s 80\% probability in Columba membership.

{\it HIP 116215}---While {\sc banyan ii} places this mid-K star in $\beta$ Pic with 77\% probability, it lacks any detectable lithium, and shows only moderately vigorous chromospheric and coronal activity.  We consider its membership to be doubtful, and provisionally assign a 20\% membership probability in $\beta$ Pic. 

\subsubsection{non-SEEDS, non-GDPS NICI MG ADI Targets}

{\it HIP 9685}---This star is a high probability (over 99\% according to {\sc banyan ii}) member of Tuc-Hor neglecting a discrepant radial velocity.  However, \cite{Biller+Liu+Wahhaj+etal_2013} reported a $\sim$0.5 $M_\odot$ stellar companion at a projected separation of $0.\!\!''18$ (9 AU), naturally explaining a 7--8 km\,s$^{-1}$ shift in the radial velocity.  We therefore regard HIP 9685 as a reliable member of Tuc-Hor.

{\it TYC 5899-26-1}---With a measured radial velocity \citep[$26.7 \pm 1.5$ km\,$s^{-1}$][]{Schlieder+Lepine+Simon_2010} and parallax \citep[$61.4 \pm 1.5$ mas,][]{Shkolnik+Anglada+Liu+etal_2012}, {\sc banyan ii} confirms AB Dor membership with 99.9\% probability.  This is consistent with the star's strong X-ray activity.

{\it HIP 26373}---{\sc Banyan} places this star in AB Dor with 100\% probability, falling to 87\% in {\sc banyan ii}.  It shows a wide range of youth indicators consistent with an age of $\sim$100 Myr, including abundant lithium, rapid rotation, and saturated chromospheric and X-ray emission.  We consider it to be a high-probability member of AB Dor.

{\it TYC 5361-1476-1}---\cite{daSilva+Torres+deLaReza+etal_2009} classified this star as a member of AB Dor, but {\sc banyan ii} places higher probability on its membership in $\beta$ Pic (84\% in $\beta$ Pic; 3\% in AB Dor, 13\% in the field).  The star does have strong youth indicators, including  vigorous X-ray activity, rapid rotation, and abundant lithium, and is very likely to be young.  Unfortunately, conclusive membership in a MG will require the measurement of a radial velocity and a parallax.  For now, we adopt a 90\% membership probability in AB Dor, effectively being conservative about its identification with $\beta$ Pic (and the younger age this would imply).

{\it TYC 7084-794-1} (= CD$-$35 2722)---While {\sc banyan ii}'s membership probability of 85\% is much lower than {\sc banyan}'s 99.9\%, the star shows strong indicators of youth, and we consider it to be a reliable member of AB Dor.  The star does have a measured radial velocity, but no parallax.

{\it HIP 30034}---\cite{Zuckerman+Song_2004} originally proposed this star to be a member of Tuc-Hor, though \cite{Torres+Quast+Melo+etal_2008} favor membership in Columba and \cite{Malo+Doyon+Lafreniere+etal_2013} favor membership in Carina.  All of these associations are of similar age ($\sim$30 Myr), making the distinction not particularly meaningful for our analysis.  HIP 30034 does show strong youth indicators and is highly likely to be young; {\sc banyan} and {\sc banyan ii} both strongly disfavors membership in the field.  We adopt a 95+\% membership probability in Carina, with the understanding that this probability encompasses HIP 30034's possible membership in Columba (or, less likely, in Tuc-Hor).  

{\it GSC 08894-00426}---This M dwarf lacks a radial velocity or parallax, and is also almost fully depleted in lithium.  {\sc Banyan} disfavors field membership, placing a 70\% probability on $\beta$ Pic membership, 12\% in Argus, and 16\% in AB Dor, while {\sc banyan ii} gives an 89\% probability of $\beta$ Pic membership and a 6\% probability of AB Dor membership, with a 4\% probability of membership in the field.  \cite{Craig+Christian+Dupuis+etal_1997} list a spectral type of M5, which would make its lack of lithium somewhat surprising in the context of likely low-mass members of $\beta$ Pic \citep{Binks+Jeffries_2014}; however, it is sufficiently close to the lithium depletion boundary to make a conclusive statement impossible.  Unfortunately, a firm membership in any MG must await a radial velocity and parallax measurement.  We provisionally (and conservatively) assign it to AB Dor, the oldest MG of which GSC 08894-00426 is a plausible member.

{\it GJ 9251 A}---Listed as a member of AB Dor by \cite{Biller+Liu+Wahhaj+etal_2013}, this star is conclusively associated with the field by {\sc banyan} ($>$99\% probability).  In addition, it has only an upper limit on X-ray luminosity, and lacks any detected rotation period or chromospheric emission; \cite{Lopez-Santiago+Montes+Galvez-Ortiz+etal_2010} measure $v \sin i < 1$ km\,s$^{-1}$.  We therefore reject the proposed MG association.

{\it TYC 8728-2262-1}---This star presents an extreme disagreement between the results using {\sc banyan}, which gives a 99\% probability of $\beta$ Pic membership, and {\sc banyan ii}, which gives the probability as just 4\%.  The star is certainly young: it shows saturated X-ray emission, rapid rotation, and abundant lithium.  Lacking a parallax, we adopt a 50\% probably of $\beta$ Pic membership.  

{\it TYC 9073-762-1}---This star is another case of a strong disagreement between {\sc banyan} (99.9\% in $\beta$ Pic) and {\sc banyan ii} (67\% in $\beta$ Pic).  As for TYC 8728-2262-1 above, the star shows abundant youth indicators but lacks a measured parallax.  We adopt an 80\% probability of $\beta$ Pic membership.

{\it TYC 7408-54-1}---This star is similar to the two cases above, with nonuniform priors reducing {\sc banyan ii}'s probability of $\beta$ Pic membership from 99.9\% to 60\%.  Due to its abundant youth indicators, we adopt an 80\% membership probability.  The measurement of a parallax will clear up each of these three cases.  

{\it HIP 95261}---This A star is known to have a brown dwarf companion \citep{Lowrance+Schneider+Kirkpatrick+etal_2000}.  Originally proposed as a member of TW Hya \citep{Zuckerman+Webb_2000} but soon reclassified into $\beta$ Pic \citep{Zuckerman+Song+Bessell+etal_2001}, {\sc banyan} rejects membership in any MG due to a discrepant radial velocity.  \cite{Malo+Doyon+Lafreniere+etal_2013} suggest that the radial velocity could be affected by the late-type companion, though a $\sim$0.2 $M_\odot$ companion 200 AU ($4''$ projected at 50 pc) from a $\sim$2 $M_\odot$ primary would induce a radial velocity of just a few tenths of a km\,s$^{-1}$, far too little to explain a $\sim$15 km\,s$^{-1}$ discrepancy.  \citeauthor{Malo+Doyon+Lafreniere+etal_2013} also suggest that, with such a rapid rotator \citep[$v \sin i = 330$ km\,s$^{-1}$,][]{daSilva+Torres+deLaReza+etal_2009}, the radial velocity could simply be in error.  We acknowledge this as a possibility, but provisionally adopt a much lower $\beta$ Pic membership probability of 50\%.

{\it 2MASS J19560294-3207186 and TYC 7443-1102-1}---These stars share a common proper motion, but lack a measured parallax.  \cite{Kiss+Moor+Szalai+etal_2011} have measured a radial velocity for TYC 7443-1102-1, resulting in a 99.8\% membership probability in $\beta$ Pic according to {\sc banyan}, and 58\% according to {\sc banyan ii}.  They show clear signs of youth, and we adopt an 80\% membership probability for the pair.

{\it HIP 99273}---This mid-F star, like the Tycho stars above, sees its membership probability in $\beta$ Pic drop from 99.9\% to $\sim$80\% from {\sc banyan} to {\sc banyan ii}.  Given the (unsurprising for its temperature) lack of secondary age indicators to confirm its youth, we adopt the latter probability.  

{\it HIP 104308}---This A star is a proposed member of TW Hya \citep{Zuckerman+Song_2004}.  \cite{Malo+Doyon+Lafreniere+etal_2013} find a 99\% probability of membership when neglecting RV data, which consist of a single uncertain measurement by \cite{Zuckerman+Song_2004}.  The addition of the radial velocity, though at $-10 \pm 10$ km\,s$^{-1}$, only 1$\sigma$ away from the TW Hya distribution of $-0.2 \pm 3.5$ km\,s$^{-1}$, reduces {\sc banyan}'s membership probability to zero (for comparison, HIP 95261 lies 5$\sigma$ from its predicted radial velocity).  We decline to draw such a strong conclusion, adopting an 80\% membership probability.

\LongTables
\onecolumngrid
\begin{deluxetable*}{lccccccr}
\tablewidth{0pt}
\tablecaption{Secondary Age Indicators}
\tablehead{
    \colhead{Name} &
    \colhead{Moving} &
    \colhead{Prob.} &
    \colhead{$\log R'_{\rm HK}$\tablenotemark{b}} &
    \colhead{$\log R_{\rm X}$\tablenotemark{c}} &
    \colhead{Li EW (m\AA)/} &
    \colhead{$P_{\rm rot}$} &
    \colhead{References\tablenotemark{e}} \\
    \colhead{HIP/HD/GJ/Other} &
    \colhead{Group} &
    \colhead{Member\tablenotemark{a}} &
    \colhead{} & 
    \colhead{} & 
    \colhead{$\log$\,A\tablenotemark{d}} &
    \colhead{(days)} &
    \colhead{}
}
    
\startdata
\cutinhead{SEEDS Debris Disk Targets}
HIP 682 & \ldots & \ldots & $-$4.38 & $-$4.15 & 150 & \ldots & T06  \\
HIP 5944 & \ldots & \ldots & $-$4.47 & $-$4.68 & [A] 2.73 & 5.67 & Mi12, G00   \\
HIP 6878 & \ldots & \ldots & $-$4.43 & $-$4.34 & 48 & 3.13 & W07, W11  \\
HIP 7345 & \ldots & \ldots & \ldots & $<$$-$5.89 & \ldots    & \ldots & \ldots   \\
HIP 7576 & $\beta$ Pic & 95$+$ & $-$4.49 & $-$4.43 & 103 & 7.15 & L06, S09   \\
HIP 9141 & Tuc-Hor & 95$+$ & $-$4.22 & $-$3.97 & 190 & 3.02 & T06, W11  \\
HIP 11360 & Columba & 85 & \ldots    & $-$4.86 & \ldots    & \ldots  & \ldots \\
HIP 11847 & $\beta$ Pic & 50 & \ldots    & $<$$-$4.69 & \ldots    & \ldots & \ldots \\
HIP 13402 & $\beta$ Pic & 20 & $-$4.36\tablenotemark{MW} & $-$3.90 & 224 & 6.76 & W07, W11  \\
HIP 18859 & AB Dor & 95$+$ & $-$4.40 & $-$4.16 & \ldots & \ldots  & \ldots \\
HD 281691 & \ldots & \ldots & $-$3.97 & $-$3.13\tablenotemark{B} & 345 & 2.65 & W07, W11   \\
HIP 22845 & \ldots & \ldots & \ldots & $<$$-$6.31 & \ldots & \ldots & \ldots \\
HIP 28103 & \ldots & \ldots & \ldots & $-$5.75 & \ldots & \ldots & \ldots  \\
HIP 37170 & \ldots & \ldots & $-$4.38 & $-$4.26 & 107 & \ldots & W07   \\
HIP 40693 & \ldots & \ldots & $-$5.03 & $-$5.90 & 2.5 & \ldots & T05   \\
HD 70573 & \ldots & \ldots & $-$4.16 & $-$4.09 & 166 & 3.30 & W07, S09   \\
HIP 42333 & \ldots & \ldots & $-$4.55 & $-$4.78\tablenotemark{B} & [A] 2.35 & 6.14 & T05, G00   \\
HIP 42430 & \ldots & \ldots & $-$5.11 & $-$6.03\tablenotemark{B} & \ldots & \ldots & \ldots  \\
HIP 42438 & UMa & 60 & $-$4.37\tablenotemark{MW} & $-$4.45 & 120 & 4.89 & W07, G00  \\
HIP 43726 & \ldots & \ldots & $-$4.64\tablenotemark{MW} & $-$5.25 & 19 & \ldots & T05  \\
HIP 49809 & \ldots & \ldots & \ldots & $-$6.22 & \ldots & \ldots & \ldots  \\
HIP 51658 & \ldots & \ldots & \ldots & $<$$-$6.27 & \ldots & \ldots & \ldots  \\
HIP 52462 & \ldots & \ldots & $-$4.40 & $-$4.37 & 138 & 13.5 & T06, S09  \\
HIP 57632 & Argus & 90 & \ldots & $<$$-$7.29 & \ldots    & \ldots & \ldots \\
HIP 58876 & \ldots & \ldots & $-$4.36 & $-$4.36 & 122 & \ldots & W07  \\
HIP 59774 & UMa & 95$+$ & \ldots & $<$$-$6.77 & \ldots & \ldots & \ldots \\
HIP 60074 & \ldots & \ldots & $-$4.33 & $-$4.33 & 123 & 7.13 & W07, K02   \\
HIP 61174 & \ldots & \ldots & \ldots & $-$5.59 & \ldots & \ldots & \ldots  \\
HIP 61498 & TW Hya & 95$+$  & \ldots & $-$5.24 & 550\tablenotemark{s} & \ldots & D09 \\
HIP 61960 & \ldots & \ldots & \ldots    & $<$$-$6.22 & \ldots    & \ldots  & \ldots \\
HIP 63076 & \ldots & \ldots & \ldots    & $<$$-$6.08 & \ldots    & \ldots & \ldots  \\
HIP 63584 & \ldots & \ldots & \ldots    & $-$5.00\tablenotemark{B} & \ldots    & \ldots & \ldots  \\
HIP 69732 & \ldots & \ldots & \ldots    & $<$$-$6.80 & \ldots    & \ldots & \ldots  \\
HIP 70952 & \ldots & \ldots & \ldots    & $-$5.03 & \ldots    & \ldots & \ldots  \\
HIP 71284 & \ldots & \ldots & \ldots    & $-$5.55 & 3 & \ldots & T05  \\
HIP 71395 & \ldots & \ldots & $-$4.52 & $-$4.47 & [A] $-0.3$ & 11.5 & M08, W11  \\
HIP 74702 & \ldots & \ldots & $-$4.50 & $<$$-$5.18 & 1.4 & 5.97 & G09, G00  \\
HIP 76267 & UMa & 95$+$ & \ldots & $-$6.64 & \ldots & \ldots & \ldots  \\
HIP 76829 & $\beta$ Pic & 84 & \ldots & $-$4.98 & \ldots    & \ldots & \ldots  \\
HIP 77542 & \ldots\tablenotemark{PMS} & \ldots & \ldots & $-$4.45\tablenotemark{B} & 500\tablenotemark{s} & \ldots & F08  \\
HIP 79977 & USco & 95$+$ & \ldots    & $<$$-$4.27 & 66 & \ldots & C11  \\
HIP 82587 & \ldots & \ldots & \ldots    & $-$5.54 & \ldots    & \ldots  & \ldots \\
HIP 87108 & \ldots & \ldots  & \ldots    & $<$$-$6.61 & \ldots    & \ldots  & \ldots \\
HIP 87558 & \ldots & \ldots  & \ldots    & $-$5.33 & \ldots    & \ldots  & \ldots \\
HIP 92919 & \ldots & \ldots & $-$4.23 & $-$3.12 & 20 & 2.91 & W07, P05 \\
HIP 95793 & \ldots & \ldots & \ldots    & $<$$-$5.80 & \ldots    & \ldots & \ldots  \\
HIP 99711 & \ldots & \ldots & $-$4.64 & $-$5.07 & [A] $-0.24$ & 24.0 & G10, S09  \\
HIP 107649 & \ldots & \ldots & $-$4.87 & $-$5.61 & 38 & \ldots & T06 \\
\cutinhead{non-SEEDS GDPS Targets}
HIP 919 & \ldots & \ldots &    $-$4.35   &    $-$4.19   &   122 &  6.05 & W07, W11 \\
HD 1405 & AB Dor & 95$+$ &   \ldots   &    $-$2.90   &   267 &  1.76 & D09, M10 \\
HIP 4907 & \ldots & \ldots &  $-$4.48 &  $-$4.73  & \ldots  &  12.2 & S09 \\
HIP 7235 & \ldots & \ldots &  $-$4.63 &  $-$5.14  & \ldots  &  \ldots & \ldots \\
HIP 9291 & Columba & 75 &  \ldots  &  $-$3.04  & \ldots  & \ldots & \ldots \\
HIP 11072 & \ldots & \ldots &  $-$5.00 &  $-$4.51  &  50  &  \ldots & T06 \\
HIP 12530 & \ldots & \ldots &  $-$4.45 &  $-$4.54  &  \ldots &  \ldots  & \ldots \\
HIP 12926 & \ldots & \ldots &  $-$4.93 &  $<$$-$5.08  & \ldots & \ldots  & \ldots \\
HIP 13081 & \ldots & \ldots &  $-$4.57  &  $-$4.47  &  \ldots  &  \ldots  & \ldots \\
HIP 14150 & \ldots & \ldots &  $-$4.86  &  $-$6.03  &  3 & \ldots & T05 \\
HIP 14954 & \ldots & \ldots &  $-$4.55  &  $<$$-$6.10  &  12 & \ldots & T05 \\
HIP 15323 & \ldots & \ldots &  $-$4.47   &  $-$4.49  &  [A] 2.67 & \ldots  & Go10 \\
BD$-$19 660 & \ldots & \ldots &   \ldots  &  $<$$-$3.53  &  \ldots  & \ldots & \ldots \\
HIP 16537 & \ldots & \ldots & $-$4.52\tablenotemark{MW} &  $-$4.68 &  1  &  11.68 & T05, W11 \\
HIP 17695 & \ldots & \ldots &   \ldots  &  $-$3.07 &  0 &   3.87  & D09, M10 \\
HIP 21482 & \ldots & \ldots & $-$4.06  &   $-$3.05  &  [A] 0.47 & \ldots & Mi12 \\
HIP 22449 & \ldots & \ldots & $-$4.71 &  $-$4.99  &  \ldots  &  \ldots & \ldots \\
HIP 23200 & $\beta$ Pic & 95$+$ &  $-$4.06  & $-$3.20  &   270 &  1.86 & T06, W11 \\
HIP 30920 & \ldots & \ldots & \ldots & $-$3.30\tablenotemark{B}  &  \ldots  &  0.39  & W11 \\
HIP 37766 & \ldots & \ldots & $-$3.62 & $-$3.00   &  \ldots & 2.78 & W11 \\
HIP 43410 & \ldots & \ldots &    $-$4.50\tablenotemark{MW} & $-$4.35 & \ldots &  \ldots & \ldots \\
HIP 44458 & \ldots & \ldots & $-$4.30 & $-$3.93\tablenotemark{B} & 170 & 2.86 & W07, W11 \\
HD 78141 & \ldots & \ldots & \ldots & $-$3.86  &   \ldots & \ldots & \ldots \\
HIP 46816 & \ldots & \ldots & $-$4.02 & $-$3.01 & 245 &  1.70 & W07, W11 \\
HIP 51386 & \ldots & \ldots &  $-$4.46 & $-$4.19  & 130 &  2.60 & W07, W11 \\
HIP 51931 & \ldots & \ldots &  $-$4.55 &   $-$4.92  &  \ldots  &  \ldots & \ldots \\
HIP 52787 & \ldots & \ldots &    $-$4.46  &  $-$4.42  &  100 &  \ldots & T06 \\
HIP 54745 & \ldots & \ldots &    $-$4.42  &  $-$4.52  &  [A] 2.55 & 7.60 & M08, W11 \\
HIP 57370 & \ldots & \ldots &    $-$4.59 & $-$5.12  &  [A] 0.46 &   12.3 & G10, W11 \\
HIP 57494 & \ldots & \ldots &  $-$4.92 &  $<$$-$4.35  &  \ldots & \ldots & \ldots \\
HD 108767B & \ldots & \ldots &   \ldots   &  $-$4.47 &  \ldots  &  \ldots & \ldots \\
BD$+$60 1417 & \ldots & \ldots &   \ldots   &  $-$4.20 &  \ldots   & \ldots & \ldots \\
HIP 62523 & \ldots & \ldots &    $-$4.62 &  $-$5.08  &  [A] 1.5  &  15.8 & M08, S09 \\
HIP 63742 & AB Dor & 95$+$ &    $-$4.43 & $-$4.26  &  142 & 6.47 & D09, W11 \\
HIP 65016 & \ldots & \ldots & $-$4.91 &   $<$$-$4.47  &  \ldots  & \ldots & \ldots \\
HIP 65515 & \ldots & \ldots &  $-$4.47 &  $-$4.36 &  [A] 1.65 &  4.27  & M08, S09 \\
HIP 67092 & \ldots & \ldots & \ldots & $<$$-$4.11  &  \ldots  &  \ldots & \ldots \\
HIP 69357 & \ldots & \ldots & $-$4.60   & $-$5.27 &   \ldots & \ldots & \ldots \\
HD 234121 & \ldots & \ldots &  $-$4.28   &  $-$4.61 & \ldots    &  \ldots & \ldots \\
HIP 71631 & \ldots & \ldots &    $-$4.19\tablenotemark{MW} & $-$3.52  &  196  &    2.67 & W07, W11 \\
HIP 72146 & \ldots & \ldots & $-$4.93   &   $<$$-$5.08 &   \ldots  &  \ldots & \ldots \\
HIP 72339 & \ldots & \ldots & $-$4.72 &  $<$$-$4.93 & [A] 0.38  &  \ldots  & G10 \\
HIP 72567 & \ldots & \ldots &  $-$4.46 & $-$4.65 & [A] 2.8 & 7.85 & M08, S09 \\
HIP 74045 & \ldots & \ldots & $-$3.99 &  $-$2.92\tablenotemark{B} & 228 & 1.90 & W07, W11 \\
HIP 75829 & \ldots & \ldots &  $-$4.40  &   $-$3.80  &  145 & \ldots & W07 \\
HIP 77408 & \ldots & \ldots &  $-$4.46 &  $-$4.41  &  [A] 0.50  &  14.05 & M08, W11 \\
HIP 79755 & \ldots & \ldots &  $-$4.92 &  $-$4.39  &  \ldots &  \ldots & \ldots \\
HIP 80824 & \ldots & \ldots & $-$5.31 &  $-$5.11  &  \ldots   &  \ldots & \ldots \\
HIP 81084 & \ldots & \ldots &    $-$4.53    &   $-$3.45  &  0 &   7.45 & D09, M10 \\
HIP 86346 & AB Dor & 95$+$ &   \ldots   &    $-$3.02 &   40 &  1.84 & D09, M10 \\
HIP 87322 & \ldots & \ldots &  \ldots  &  $<$$-$4.40  &  \ldots    &  \ldots & \ldots \\
HIP 88848 & \ldots & \ldots & $-$4.14  & $-$2.98  &  179 & \ldots & W07 \\
HIP 89005 & \ldots & \ldots &  \ldots  &  $-$3.95  &  11  &  \ldots & G09 \\
HIP 97438 & \ldots & \ldots & $-$3.77  & $-$4.69  &  87 &  \ldots & G09 \\
HIP 101262 & \ldots & \ldots &  \ldots  &  $-$4.99 &  \ldots  &  \ldots & \ldots \\
HIP 104225 & \ldots & \ldots & $-$4.97  &  $<$$-$5.24 &  \ldots  &  \ldots & \ldots \\
HIP 105038 & \ldots & \ldots & $-$4.56  & $-$5.09  & [A] $-0.75$ & \ldots & M08 \\
HIP 106231 & AB Dor & 95$+$ &  \ldots  & $-$3.09 &  215 &  0.42  & D09, M10 \\
HIP 108156 & \ldots & \ldots & $-$4.90 &  $-$5.23  & [A] 0 &  4.51 & M08, S09 \\
V383 Lac & \ldots & \ldots &  \ldots  & $-$3.06  & 270  &  2.47 & W07, W11 \\
HIP 112909 & \ldots & \ldots &  \ldots  &  $-$3.06  &  \ldots  &  1.64  & N07 \\
HIP 113020 & \ldots & \ldots &  \ldots  &  $<$$-$4.16  &  \ldots &  \ldots & \ldots \\
HIP 115147 & Columba & 80 & $-$4.16 & $-$3.15 & [A] 2.3 &  \ldots & M08 \\
HIP 116215 & $\beta$ Pic & 20 & $-$4.47 &  $-$4.51 &  0 & \ldots & T06 \\
HIP 116384 & \ldots & \ldots &  \ldots &  $-$3.60 &  15  & \ldots & T06 \\
HIP 117410 & \ldots & \ldots &  \ldots &  $-$3.47 &  0 &  \ldots & T06 \\
\cutinhead{non-SEEDS, non-GDPS NICI Moving Group ADI Targets}
HIP 560  &  $\beta$ Pic & 95+ &  \ldots  & $-$5.27  & 87 & \ldots & T06 \\
HIP 1481  &  Tuc-Hor & 95+ &   $-$4.37  & $-$4.12  & 130 & \ldots & D09 \\
HIP 2729  &  Tuc-Hor & 95+ &  \ldots  & $-$3.23  & 360 & 0.37 & D09, M10 \\
HIP 5191  &  AB Dor & 95+ &  \ldots  & $-$3.68  & 155 & 7.13 & D09, M10 \\
HIP 9685  &  Tuc-Hor & 95+ &  \ldots  & $-$4.92  & 60 & \ldots & T06 \\
HIP 12394  &  Tuc-Hor & 95+ &  \ldots  & $<$$-$6.42  & 0 & \ldots & D09 \\
TYC 8491-656-1  &  Tuc-Hor & 95+ &  \ldots  & $-$2.99  & 298 & 1.28 & D09, M10 \\
AF Hor  &  Tuc-Hor & 95+ &  \ldots  & $-$2.59  & 10 & 0.56 & D09, K12 \\
TYC 8497-995-1  &  Tuc-Hor & 95+ &  \ldots  & $-$3.55  & 120 & 7.4 & D09, M10 \\
HIP 14684  &  AB Dor & 95+ &   $-$4.38  & $<$$-$4.27  & 191 & 5.5 & D09, M10 \\
TYC 5899-26-1  &  AB Dor & 95+ &  \ldots  & $-$3.10  & 20 & \ldots & D09 \\
TYC 8513-952-3  &  AB Dor & 95+ &  \ldots  & $-$2.92  & 0 & 1.5 & D09, M10 \\
HIP 23309  &  $\beta$ Pic & 95+ &  \ldots  & $-$3.42  & 294 & 8.6 & F08, M10 \\
HIP 25283  &  AB Dor & 95+ &   $-$4.32  & $-$3.64  & 15 & 9.3 & D09, M10 \\
HIP 26373  &  AB Dor & 95+ &   $-$4.21  & $-$3.17  & 285 & 4.5 & D09, M10 \\
TYC 5361-1476-1  &  AB Dor & 90 &  \ldots  & $-$3.50  & 230 & 5.6 & D09, K12 \\
TYC 7084-794-1  &  AB Dor & 95+ &  \ldots  & $-$3.12  & 10 & 1.7 & D09, M10 \\
HIP 29964  &  $\beta$ Pic & 95+ &   $-$4.16  & $-$2.83  & 357 & 2.7 & F08, M10 \\
HIP 30034  &  Car & 95+ &  \ldots  & $-$3.16  & 320 & 3.9 & D09, M10 \\
HIP 30314  &  AB Dor & 95+ &   $-$4.34  & $-$4.14  & 150 & 0.48 & D09, R12 \\
GSC 08894-00426  &  AB Dor & 95$+$ &  \ldots  & $-$2.84  & 0 & 1.0 & D09, K12 \\
GJ 9251 A  &  AB Dor & \ldots &  \ldots  & $<$$-$4.07  & \ldots & \ldots & \ldots \\
TWA 6  &  TW Hya & 95+ &  \ldots  & $-$2.88  & 560 & 0.54 & F08, M10 \\
TWA 7  &  TW Hya & 95+ &  \ldots  & $-$3.19  & 530 & 5.0 & T06, M10 \\
HIP 53911  &  TW Hya & 95+ &  \ldots  & $-$2.12  & 426 & 2.8 & F08, M10 \\
TWA 14  &  TW Hya & 95+ &  \ldots  & $-$2.89  & 600 & 0.63 & F08, M10 \\
TWA 13N  &  TW Hya & 95+ &  \ldots  & $-$2.82  & 570 & 5.6 & F08, M10 \\
TWA 8A  &  TW Hya & 95+ &  \ldots  & $-$2.92  & 530 & 4.7 & F08, M10 \\
TWA 9B  &  TW Hya & 95+ &  \ldots  & $-$2.30  & 480 & 4.0 & D09, M10 \\
HIP 57589  &  TW Hya & 95+ &  \ldots  & $-$2.88  & 470 & 5.0 & F08, M10 \\
TWA 25  &  TW Hya & 95+ &  \ldots  & $-$3.05  & 494 & 5.1 & F08, M10 \\
TWA 20  &  TW Hya & 95+ &  \ldots  & $-$3.37  & 160 & 0.65 & F08, M11 \\
TWA 10  &  TW Hya & 95+ &  \ldots  & $-$3.02  & 460 & 8.4 & F08, M10 \\
TWA 11B  &  TW Hya & 95+ &  \ldots  & $-$3.12  & 550 & \ldots & D09 \\
TWA 11A  &  TW Hya & 95+ &  \ldots  & $-$5.27  & 550 & \ldots & D09 \\
HD 139084B  &  $\beta$ Pic & 95+ &  \ldots  & $-$1.58\tablenotemark{B}  & 260 & 4.3 & F08, M10 \\
HD 155555C  &  $\beta$ Pic & 95+ &  \ldots  & $-$1.67\tablenotemark{B}  & 250 & 1.7 & D09, M10 \\
TYC 8728-2262-1  &  $\beta$ Pic & 50 &  \ldots  & $-$3.08  & 360 & 1.8 & D09, M10 \\
HD 164249B  &  $\beta$ Pic & 95+ &  \ldots  & $-$3.01\tablenotemark{B}  & 92 & \ldots & F08 \\
HIP 92024  &  $\beta$ Pic & 95+ &  \ldots  & $-$5.64  & 490 & 0.35 & T06, M10 \\
TYC 9073-762-1  &  $\beta$ Pic & 80 &  \ldots  & $-$3.21  & 332 & 5.4 & D09, M10 \\
TYC 7408-54-1  &  $\beta$ Pic & 80 &  \ldots  & $-$3.04  & 492 & 1.1 & D09, M10 \\
HIP 92680  &  $\beta$ Pic & 95+ &   $-$3.82  & $-$3.13  & 287 & 1.0 & D09, M10 \\
HIP 95261  &  $\beta$ Pic & 50 &  \ldots  & $<$$-$5.73  & \ldots & \ldots & \ldots \\
HIP 95270  &  $\beta$ Pic & 95+ &  \ldots  & $<$$-$4.39  & 120 & \ldots & D09 \\
2MASS J19560294-3207186  &  $\beta$ Pic & 80 &  \ldots  & $-$2.85\tablenotemark{B}  & 500 & \ldots & Mc12 \\
TYC 7443-1102-1  &  $\beta$ Pic & 80 &  \ldots  & $-$3.00\tablenotemark{B}  & 110 & 12 & M11, Mc12 \\
HIP 99273  &  $\beta$ Pic & 80 &   $-$4.59  & $-$4.83  & 58 & \ldots & W07 \\
HIP 104308  &  Tuc-Hor & 80 &  \ldots  & $<$$-$5.34  & \ldots & \ldots & \ldots \\
HIP 107345  &  Tuc-Hor & 95+ &  \ldots  & $-$3.22  & 55 & 4.5 & D09, M10 \\
TYC 9340-437-1  &  $\beta$ Pic & 95+ &  \ldots  & $-$2.90  & 440 & 4.5 & D09, M10 \\
HIP 112312  &  $\beta$ Pic & 95+ &  \ldots  & $-$2.68\tablenotemark{B}  & 0 & 2.4 & D09, M10 \\
TX PsA  &  $\beta$ Pic & 95+ &  \ldots  & $-$2.31\tablenotemark{B}  & 450 & \ldots & D09 \\
TYC 5832-666-1  &  $\beta$ Pic & 95+ &  \ldots  & $-$3.05  & 185 & 5.7 & D09, M10 \\
HIP 118121  &  Tuc-Hor & 95+ &  \ldots  & $-$5.79  & \ldots & \ldots & \ldots \\
\cutinhead{Additional HiCIAO Detections}
HIP 64792 (GJ 504) & \ldots & \ldots & $-$4.45\tablenotemark{MW} & $-$4.42 & 83 & 3.3 & T05, M03 \\
HIP 95319 (GJ 758) & \ldots & \ldots & $-$5.08 & $<$$-$5.04 & 2 & \ldots & T05 
\enddata

\tablenotetext{a}{High-probability (95\%$+$) members typically have a 99\%$+$ membership probability from {\sc Banyan} \citep{Malo+Doyon+Lafreniere+etal_2013}, or a 95\%+ probability from {\sc Banyan ii} \citep{Gagne+Lafreniere+Doyon+etal_2014}, plus an additional youth indicator.  See Section \ref{subsec:uncertain_mg} for details on other stars.}
\tablenotetext{b}{From the catalog compiled by \cite{Pace_2013}; see text for details.  All targets bluer than $B-V=0.45$ have been omitted.}
\tablenotetext{c}{From the {\it ROSAT} satellite; see text for details.}
\tablenotetext{d}{Logarithmic lithium abundances, ${\rm A(H)} = 12$, are preceded by [A].  The other values are equivalent widths.}
\tablenotetext{e}{References abbreviated as: C11 \citep{Chen+Mamajek+Bitner+etal_2011}; D09 \citep{daSilva+Torres+deLaReza+etal_2009}; F08 \citep{Fernandez+Figueras+Torra_2008}; G00 \citep{Gaidos+Henry+Henry_2000}; G09 \citep{Guillout+Klutsch+Frasca+etal_2009}; G10 \citep{Ghezzi+Cunha+Smith+etal_2010}; K02 \citep{Koen+Eyer_2002}; K12 \citep{Kiraga_2012}; L06 \citep{Lopez-Santiago+Montes+Crespo-Chacon+etal_2006}; M96 \citep{Messina+Guinan_1996}; M03 \citep{Messina+Pizzolato+Guinan+etal_2003}; M08 \citep{Mishenina+Soubiran+Bienayme+etal_2008}; M10 \citep{Messina+Desidera+Tutatto+etal_2010}; M11 \citep{Messina+Desidera+Lanzafame+etal_2011}; Mi12 \citep{Mishenina+Soubiran+Kovtyukh+etal_2012}; Mc12 \citep{McCarthy+White_2012}; N07 \citep{Norton+Wheatley+West+etal_2007}; P05 \citep{Pojmanski+Pilecki+Szczygiel_2005}; S09 \citep{Samus+Durlevich+etal_2009}; T06 \citep{Torres+Quast+daSilva+etal_2006}, T05 \citep{Takeda+Kawanomoto_2005}; W07 \citep{White+Gabor+Hillenbrand_2007}; W11 \citep{Wright+Drake+Mamajek+etal_2011}}
\tablenotetext{s}{Equivalent width in the secondary}
\tablenotetext{B}{The star has a known binary companion that could contribute X-ray flux.}
\tablenotetext{MW}{$R'_{\rm HK}$ from multi-decade Mt.~Wilson measurements \citep{Baliunas+Donahue+Soon+etal_1995, Radick+Lockwood+Skiff+etal_1998, Lockwood+Skiff+Henry+etal_2007}}
\tablenotetext{PMS}{Pre-main-sequence object with an age of $5 \pm 1$ Myr}

\label{tab:age_indicators}

\end{deluxetable*}
\twocolumngrid

\section{Bayesian Ages} \label{sec:bayes_ages}

We revisit the age estimates of all targets in our sample using the method detailed in \cite{Brandt+Kuzuhara+McElwain+etal_2014}, which we summarize here.  Our method combines a possible identification with a MG of known age with secondary age indicators, including chromospheric and X-ray activity and stellar rotation.  We use the MGs identified in that paper, to which we refer for discussion and references:
\begin{enumerate}
\item $\beta$ Pictoris ($21 \pm 4$ Myr)
\item AB Doradus ($130 \pm 20$ Myr)
\item Columba ($30^{+20}_{-10}$ Myr)
\item Tucana-Horologium ($30^{+10}_{-20}$ Myr)
\item TW Hydrae ($10 \pm 5$ Myr)
\item Ursa Major ($500 \pm 100$ Myr)
\end{enumerate}
We treat all age uncertainties as $2\sigma$ limits, and approximate the age probability distribution for each group as a Gaussian.  Our sample also contains one star each from Upper Scorpius, Argus, and Carina. Upper Scorpius is a nearby very young ($\sim$10 Myr) starforming region \citep{deZeeuw+Hoogerwerf+deBruijne+etal_1999}, while Argus and Carina have similar ages to Columba \citep[$\sim$30 Myr, ][]{Torres+Quast+Melo+etal_2008}.

We estimate the membership probability $p_{\rm MG}$ for each star as described in Section \ref{subsec:uncertain_mg}, basing our own probabilities primarily on the {\sc Banyan} and {\sc Banyan ii} Bayesian analyses, and capping the membership probability at 95\%.  The probability distribution ${\cal P}(\tau)$ for the stellar age is then
\begin{equation}
{\cal P}(\tau) = p_{\rm MG} {\cal P}_{\rm MG} (\tau) + \left(1 - p_{\rm MG} \right) {\cal P}(\tau|{\rm indic} )~,
\label{eq:age_dist}
\end{equation}
where ${\cal P}_{\rm MG}$ is the probability distribution for the age of the MG and ${\cal P}(\tau|{\rm indic} )$ is the probability distribution for the age given the observed activity and rotation indicators.  We calculate this probability distribution as described in \cite{Brandt+Kuzuhara+McElwain+etal_2014}; we summarize the method here.

Main-sequence stars cooler than spectral type late-F have large convective zones in their outer regions.  These regions support magnetic dynamos powered by the stellar differential rotation, which drive chromospheric and coronal activity and magnetized stellar winds.  Over time, the stellar wind carries away angular momentum, and the star spins down.  Stellar rotation, X-ray, and chromospheric activity in the Ca\,{\sc ii} HK line all therefore decline with time, and may be calibrated as crude clocks.  These calibrations depend on the properties of the convective zone, generally parametrized using $B-V$ color for main-sequence stars.  Young stars also spend a variable amount of time (longer for cooler stars) on a rapidly rotating $C$-sequence before approaching solid-body rotation and beginning to spin down.

\cite{Mamajek+Hillenbrand_2008} have calibrated each of these secondary indicators, $R_{\rm X}$ (the ratio of X-ray to bolometric power), $R'_{\rm HK}$ (the ratio of energy in Ca\,{\sc ii} HK emission to the underlying photospheric continuum), and rotation period, using a large sample of Solar-type stars.  We use each of these calibrations, with a correction for the time a star takes to settle on the slowly rotating $I$-sequence, as described in \cite{Brandt+Kuzuhara+McElwain+etal_2014} and references therein.  We do not assume the chromospheric and coronal activity measurements to be independent age indicators, but rather treat them both as proxies for the Rossby number (the ratio of the rotation period to the convective overturn timescale $\tau_C$), which we convert to an age using a color-dependent estimate of $\tau_C$ and the period-age relations.  

We then combine the activity age distribution with that inferred directly from the rotation period (if measured) and add an additional 15\% (0.06 dex) to account for systematic uncertainties.  We also add a 5\% probability that the star is a pathological case, in the sense that its secondary age indicators do not correspond to its actual age and are useless for dating.  Multiplying the resulting distribution by a prior uniform in time out to either 10 Gyr or the star's color-dependent main sequence lifespan gives ${\cal P}(\tau|{\rm indic})$.  Finally, we weight ${\cal P}(\tau|{\rm indic})$ by $1-p_{\rm MG}$, the probability that the star is {\it not} a member of a well-defined MG, and use it in Equation \eqref{eq:age_dist}.

When drawing ages for each star from these probability distributions, we account for the fact that the ages of each member of a MG should be the same.  We 
first draw the ages for each of the MGs, then assign each suggested group member either to its group or to the field, and finally draw an age for each field star from its posterior age probability distribution.  Given the large number of stars, the central limit theorem drives the predicted number of substellar companions to a relatively narrow Gaussian given a fixed substellar distribution function.

Table \ref{tab:allages} lists the ages (in Myr) at 5\%, 10\%, 25\%, 50\%, 75\%, 90\%, and 95\% of the cumulative posterior probability.  The range of ages between the 25\% and 75\% levels thus contains half of the posterior probability, while the range from 5\% to 95\% contains nine-tenths of the posterior probability, and may be used as a 90\% confidence age interval.  We use the full posterior probability distributions summarized in Table \ref{tab:allages} throughout the rest of our analysis.  A few of these differ significantly from the distributions in \cite{Brandt+Kuzuhara+McElwain+etal_2014}; these targets had their Ca\,{\sc ii} HK and/or X-ray activity incorrectly input into the earlier paper's calculation.

\LongTables
\begin{deluxetable*}{lccccccccr}
\tablewidth{0pt}
\tablecaption{Age Distributions of All Targets}
\tablehead{
    \colhead{Name} &
    \colhead{RA (J2000)} &
    \colhead{Dec (J2000)} &
    \multicolumn{7}{c}{Age at Posterior CDF Value (Myr)} \\
    \colhead{HIP/HD/GJ/Other} &
    \colhead{(h m s)} &
    \colhead{($^\circ$ $'$ $''$)} &
    \colhead{5\%} & 
    \colhead{10\%} & 
    \colhead{25\%} & 
    \colhead{50\%} & 
    \colhead{75\%} & 
    \colhead{90\%} & 
    \colhead{95\%} 
}
\startdata 
 HIP 544 & 00 06 36.8 &  $+$29 01 17 & 192 & 208 & 237 & 272 & 311 & 356 & 421 \\
 HIP 560  & 00 06 50.1 &  $-$23 06 27 & 17 & 18 & 19 & 21 & 22 & 24 & 27 \\
 HIP 682 & 00 08 25.7 &  $+$06 37 00 & 21 & 43 & 108 & 217 & 343 & 498 & 725 \\
 HIP 919 & 00 11 22.4 &  $+$30 26 58 & 205 & 221 & 251 & 287 & 326 & 372 & 436 \\
 HIP 1134 & 00 14 10.3 &  $-$07 11 57 & 21 & 23 & 26 & 30 & 37 & 45 & 56 \\
 HD 1405 & 00 18 20.9 &  $+$30 57 22 & 109 & 115 & 122 & 129 & 136 & 142 & 146 \\
 HIP 1481  & 00 18 26.1 &  $-$63 28 39 & 13 & 17 & 23 & 30 & 33 & 37 & 41 \\
 FK Psc & 00 23 34.7 &  $+$20 14 29 & 19 & 20 & 194 & 300 & 389 & 455 & 510 \\
 HIP 2729  & 00 34 51.2 &  $-$61 54 58 & 13 & 17 & 23 & 30 & 33 & 37 & 40 \\
 HIP 3589 & 00 45 50.9 &  $+$54 58 40 & 111 & 115 & 122 & 129 & 137 & 143 & 148 \\
 HIP 4907 & 01 02 57.2 &  $+$69 13 37 & 686 & 732 & 812 & 912 & 1030 & 1160 & 1320 \\
 HIP 4979 & 01 03 49.0 &  $+$01 22 01 & 152 & 304 & 766 & 1530 & 2300 & 2760 & 2920 \\
 HIP 5191  & 01 06 26.2 &  $-$14 17 47 & 113 & 117 & 123 & 130 & 138 & 146 & 168 \\
 HIP 5944 & 01 16 29.3 &  $+$42 56 22 & 323 & 346 & 388 & 440 & 500 & 572 & 668 \\
 HIP 6869 & 01 28 24.4 &  $+$17 04 45 & 787 & 876 & 1040 & 1240 & 1510 & 1880 & 2500 \\
 HIP 6878 & 01 28 34.4 &  $+$42 16 04 & 226 & 255 & 304 & 367 & 458 & 660 & 2870 \\
 HIP 7235 & 01 33 15.8 &  $-$24 10 41 & 1340 & 1500 & 1810 & 2200 & 2680 & 3240 & 3740 \\
 HIP 7345 & 01 34 37.8 &  $-$15 40 35 & 62 & 124 & 313 & 625 & 938 & 1130 & 1200 \\
 HS Psc & 01 37 23.2 &  $+$26 57 12 & 111 & 116 & 122 & 130 & 137 & 144 & 149 \\
 HIP 7576 & 01 37 35.5 &  $-$06 45 38 & 17 & 18 & 19 & 21 & 22 & 24 & 29 \\
 HIP 9141 & 01 57 49.0 &  $-$21 54 05 & 13 & 17 & 23 & 30 & 34 & 38 & 50 \\
 HIP 9291 & 01 59 23.5 &  $+$58 31 16 & 21 & 23 & 27 & 33 & 47 & 304 & 561 \\
 HIP 9685  & 02 04 35.1 &  $-$54 52 54 & 13 & 17 & 23 & 30 & 34 & 38 & 46 \\
 HIP 10679 & 02 17 24.7 &  $+$28 44 30 & 17 & 18 & 19 & 21 & 22 & 24 & 26 \\
 HIP 11072 & 02 22 32.5 &  $-$23 48 59 & 1820 & 2000 & 2340 & 2770 & 3280 & 3860 & 4330 \\
 HIP 11360 & 02 26 16.2 &  $+$06 17 33 & 22 & 24 & 27 & 32 & 41 & 1610 & 3220 \\
 BD$+$30 397B & 02 27 28.0 &  $+$30 58 41 & 17 & 18 & 19 & 21 & 22 & 24 & 26 \\
 HIP 11437 & 02 27 29.3 &  $+$30 58 25 & 17 & 18 & 19 & 21 & 22 & 24 & 29 \\
 HIP 11847 & 02 32 55.8 &  $+$37 20 01 & 18 & 19 & 20 & 25 & 2120 & 3400 & 3820 \\
 HIP 12394  & 02 39 35.4 &  $-$68 16 01 & 13 & 17 & 23 & 30 & 33 & 37 & 43 \\
 HIP 12530 & 02 41 14.0 &  $-$00 41 44 & 188 & 267 & 421 & 1310 & 3730 & 5180 & 5670 \\
 HIP 12545 & 02 41 25.9 &  $+$05 59 18 & 17 & 18 & 19 & 21 & 22 & 24 & 25 \\
 TYC 8491-656-1  & 02 41 46.8 &  $-$52 59 52 & 13 & 17 & 23 & 30 & 33 & 37 & 40 \\
 AF Hor  & 02 41 47.3 &  $-$52 59 31 & 13 & 17 & 23 & 30 & 33 & 37 & 41 \\
 HIP 12638 & 02 42 21.3 &  $+$38 37 07 & 113 & 117 & 123 & 130 & 138 & 146 & 169 \\
 TYC 8497-995-1  & 02 42 33.0 &  $-$57 39 37 & 13 & 17 & 23 & 30 & 34 & 38 & 51 \\
 HIP 12925 & 02 46 14.6 &  $+$05 35 33 & 13 & 17 & 23 & 30 & 33 & 37 & 42 \\
 HIP 12926 & 02 46 15.2 &  $+$25 39 00 & 4260 & 4760 & 5560 & 6570 & 7770 & 8950 & 9470 \\
 HIP 13081 & 02 48 09.1 &  $+$27 04 07 & 810 & 937 & 1170 & 1480 & 1860 & 2320 & 2780 \\
 HIP 13402 & 02 52 32.1 &  $-$12 46 11 & 19 & 20 & 218 & 284 & 333 & 380 & 425 \\
 HIP 14150 & 03 02 26.0 &  $+$26 36 33 & 2920 & 3230 & 3760 & 4420 & 5200 & 6040 & 6670 \\
 HIP 14684  & 03 09 42.3 &  $-$09 34 47 & 113 & 117 & 123 & 130 & 138 & 146 & 161 \\
 HIP 14954 & 03 12 46.4 &  $-$01 11 46 & 371 & 433 & 550 & 708 & 905 & 1160 & 1440 \\
 HIP 15323 & 03 17 40.0 &  $+$31 07 37 & 179 & 233 & 314 & 416 & 544 & 708 & 928 \\
 BD$-$19 660 & 03 20 50.7 &  $-$19 16 09 & 820 & 1310 & 2770 & 5180 & 7590 & 9040 & 9510 \\
 HIP 16537 & 03 32 55.8 &  $-$09 27 30 & 619 & 659 & 728 & 812 & 909 & 1020 & 1160 \\
 HIP 17248 & 03 41 37.3 &  $+$55 13 07 & 21 & 23 & 26 & 30 & 37 & 45 & 54 \\
 HD 23061 & 03 42 55.1 &  $+$24 29 35 & 116 & 118 & 121 & 124 & 128 & 131 & 133 \\
 TYC 1803-1406-1 & 03 43 27.1 &  $+$25 23 15 & 116 & 118 & 121 & 124 & 128 & 131 & 133 \\
 HD 23247 & 03 44 23.5 &  $+$24 07 58 & 116 & 118 & 121 & 124 & 128 & 131 & 133 \\
 BD$+$23 514 & 03 45 41.9 &  $+$24 25 53 & 116 & 118 & 121 & 124 & 128 & 131 & 133 \\
 V1171 Tau & 03 46 28.4 &  $+$24 26 02 & 116 & 118 & 121 & 124 & 128 & 131 & 133 \\
 HD 23514 & 03 46 38.4 &  $+$22 55 11 & 116 & 118 & 121 & 124 & 128 & 131 & 133 \\
 HD 282954 & 03 46 38.8 &  $+$24 57 35 & 116 & 118 & 121 & 124 & 128 & 131 & 133 \\
 {\sc H\,ii} 1348 & 03 47 18.1 &  $+$24 23 27 & 116 & 118 & 121 & 124 & 128 & 131 & 133 \\
 HIP 17695 & 03 47 23.3 &  $-$01 58 20 & 67 & 84 & 132 & 211 & 291 & 339 & 365 \\
 HD 23863 & 03 49 12.2 &  $+$23 53 12 & 116 & 118 & 121 & 124 & 128 & 131 & 133 \\
 {\sc H\,ii} 2311 & 03 49 28.7 &  $+$23 42 44 & 116 & 118 & 121 & 124 & 128 & 131 & 133 \\
 HD 23912 & 03 49 32.7 &  $+$23 22 49 & 116 & 118 & 121 & 124 & 128 & 131 & 133 \\
 {\sc H\,ii} 2366 & 03 49 36.5 &  $+$24 17 46 & 116 & 118 & 121 & 124 & 128 & 131 & 133 \\
 {\sc H\,ii} 2462 & 03 49 50.4 &  $+$23 42 20 & 116 & 118 & 121 & 124 & 128 & 131 & 133 \\
 BD$+$22 574 & 03 49 56.5 &  $+$23 13 07 & 116 & 118 & 121 & 124 & 128 & 131 & 133 \\
 V1174 Tau & 03 50 34.6 &  $+$24 30 28 & 116 & 118 & 121 & 124 & 128 & 131 & 133 \\
 HIP 18050 & 03 51 27.2 &  $+$24 31 07 & 116 & 118 & 121 & 124 & 128 & 131 & 133 \\
 V1054 Tau & 03 51 39.3 &  $+$24 32 56 & 116 & 118 & 121 & 124 & 128 & 131 & 133 \\
 V885 Tau & 03 53 45.3 &  $+$25 55 34 & 116 & 118 & 121 & 124 & 128 & 131 & 133 \\
 HIP 18859 & 04 02 36.7 &  $-$00 16 08 & 112 & 116 & 123 & 130 & 137 & 144 & 150 \\
 HD 281691 & 04 09 09.7 &  $+$29 01 30 & 52 & 61 & 80 & 107 & 134 & 159 & 225 \\
 HIP 21482 & 04 36 48.2 &  $+$27 07 56 & 22 & 46 & 116 & 234 & 358 & 477 & 662 \\
 HIP 22449 & 04 49 50.4 &  $+$06 57 41 & 315 & 630 & 1400 & 2770 & 4340 & 5280 & 5600 \\
 TYC 5899-26-1  & 04 52 24.4 &  $-$16 49 22 & 112 & 116 & 123 & 130 & 137 & 145 & 152 \\
 TYC 8513-952-3  & 04 53 31.2 &  $-$55 51 37 & 110 & 115 & 122 & 129 & 136 & 143 & 148 \\
 HIP 22845 & 04 54 53.7 &  $+$10 09 03 & 67 & 136 & 339 & 679 & 1020 & 1220 & 1300 \\
 HIP 23200 & 04 59 34.8 &  $+$01 47 01 & 17 & 18 & 19 & 21 & 22 & 24 & 26 \\
 HIP 23309  & 05 00 47.1 &  $-$57 15 25 & 17 & 18 & 19 & 21 & 22 & 24 & 29 \\
 HIP 23362 & 05 01 25.6 &  $-$20 03 07 & 21 & 23 & 26 & 30 & 37 & 45 & 55 \\
 HIP 25283  & 05 24 30.2 &  $-$38 58 11 & 113 & 117 & 123 & 130 & 138 & 146 & 169 \\
 HIP 25486 & 05 27 04.8 &  $-$11 54 03 & 17 & 18 & 19 & 21 & 22 & 24 & 25 \\
 HD 36869 & 05 34 09.2 &  $-$15 17 03 & 21 & 23 & 26 & 30 & 37 & 45 & 51 \\
 HIP 26373  & 05 36 56.9 &  $-$47 57 53 & 113 & 117 & 123 & 130 & 137 & 145 & 152 \\
 HIP 28103 & 05 56 24.3 &  $-$14 10 04 & 197 & 400 & 1000 & 2010 & 3010 & 3610 & 3820 \\
 TYC 5361-1476-1  & 06 02 21.9 &  $-$13 55 33 & 113 & 117 & 124 & 131 & 139 & 152 & 232 \\
 HIP 29067 & 06 07 55.3 &  $+$67 58 37 & 96 & 193 & 439 & 673 & 941 & 1280 & 1690 \\
 TYC 7084-794-1  & 06 09 19.2 &  $-$35 49 31 & 110 & 115 & 122 & 129 & 137 & 143 & 148 \\
 HIP 29964  & 06 18 28.2 &  $-$72 02 41 & 17 & 18 & 19 & 21 & 22 & 24 & 28 \\
 HIP 30030 & 06 19 08.1 &  $-$03 26 20 & 21 & 23 & 26 & 30 & 37 & 45 & 51 \\
 HIP 30034  & 06 19 12.9 &  $-$58 03 16 & 30 & 32 & 36 & 40 & 44 & 49 & 62 \\
 HIP 30314  & 06 22 30.9 &  $-$60 13 07 & 102 & 114 & 122 & 129 & 136 & 142 & 146 \\
 GSC 08894-00426  & 06 25 56.1 &  $-$60 03 27 & 110 & 115 & 122 & 129 & 136 & 143 & 147 \\
 HIP 30920 & 06 29 23.4 &  $-$02 48 50 & 8 & 18 & 50 & 112 & 195 & 277 & 339 \\
 HIP 32104 & 06 42 24.3 &  $+$17 38 43 & 21 & 23 & 26 & 30 & 37 & 46 & 58 \\
 V429 Gem & 07 23 43.6 &  $+$20 24 59 & 111 & 116 & 122 & 130 & 137 & 143 & 148 \\
 HIP 37170 & 07 38 16.4 &  $+$47 44 55 & 24 & 50 & 127 & 255 & 393 & 562 & 806 \\
 HIP 37288 & 07 39 23.0 &  $+$02 11 01 & 1580 & 1780 & 2150 & 2620 & 3180 & 3830 & 4390 \\
 HIP 37766 & 07 44 40.2 &  $+$03 33 09 & 45 & 61 & 106 & 181 & 256 & 300 & 331 \\
 GJ 9251A  & 08 07 09.1 &  $+$07 23 00 & 840 & 1340 & 2800 & 5190 & 7600 & 9030 & 9520 \\
 FP Cnc & 08 08 56.4 &  $+$32 49 11 & 23 & 25 & 29 & 50 & 169 & 243 & 267 \\
 HIP 40693 & 08 18 23.9 &  $-$12 37 56 & 4360 & 4840 & 5630 & 6620 & 7840 & 9020 & 9510 \\
 HIP 40774 & 08 19 19.1 &  $+$01 20 20 & 181 & 355 & 584 & 840 & 1160 & 1570 & 2030 \\
 HD 70573 & 08 22 50.0 &  $+$01 51 34 & 107 & 118 & 140 & 168 & 197 & 226 & 266 \\
 HIP 42333 & 08 37 50.3 &  $-$06 48 25 & 332 & 355 & 399 & 451 & 512 & 585 & 684 \\
 HIP 42430 & 08 39 07.9 &  $-$22 39 43 & 4520 & 5020 & 5850 & 6900 & 8190 & 9260 & 9640 \\
 HIP 42438 & 08 39 11.7 &  $+$65 01 15 & 238 & 261 & 315 & 457 & 514 & 554 & 580 \\
 HIP 43410 & 08 50 32.2 &  $+$33 17 06 & 267 & 299 & 359 & 437 & 532 & 654 & 833 \\
 HIP 43726 & 08 54 17.9 &  $-$05 26 04 & 1220 & 1340 & 1560 & 1840 & 2170 & 2570 & 2960 \\
 HIP 44458 & 09 03 27.1 &  $+$37 50 28 & 100 & 111 & 132 & 160 & 188 & 215 & 250 \\
 HIP 44526 & 09 04 20.7 &  $-$15 54 51 & 282 & 305 & 346 & 395 & 448 & 508 & 589 \\
 HD 78141 & 09 07 18.1 &  $+$22 52 22 & 26 & 54 & 139 & 279 & 448 & 810 & 1370 \\
 HIP 45383 & 09 14 53.7 &  $+$04 26 34 & 41 & 83 & 207 & 409 & 620 & 872 & 1210 \\
 HIP 46816 & 09 32 25.6 &  $-$11 11 05 & 25 & 32 & 52 & 84 & 117 & 137 & 148 \\
 HIP 46843 & 09 32 43.8 &  $+$26 59 19 & 164 & 178 & 204 & 236 & 270 & 308 & 362 \\
 HIP 49809 & 10 10 05.9 &  $-$12 48 57 & 207 & 418 & 1050 & 2100 & 3150 & 3780 & 4000 \\
 HIP 50156 & 10 14 19.2 &  $+$21 04 30 & 22 & 24 & 27 & 33 & 45 & 341 & 405 \\
 TWA 6  & 10 18 28.7 &  $-$31 50 03 & 5 & 6 & 8 & 10 & 11 & 13 & 16 \\
 GJ 388 & 10 19 36.3 &  $+$19 52 12 & 41 & 56 & 104 & 182 & 261 & 308 & 327 \\
 HIP 50660 & 10 20 45.9 &  $+$32 23 54 & 1170 & 1350 & 1680 & 2110 & 2640 & 3260 & 3830 \\
 HIP 51317 & 10 28 55.6 &  $+$00 50 28 & 113 & 117 & 123 & 130 & 138 & 146 & 169 \\
 HIP 51386 & 10 29 42.2 &  $+$01 29 28 & 100 & 112 & 135 & 164 & 195 & 228 & 277 \\
 HIP 51658 & 10 33 13.9 &  $+$40 25 32 & 126 & 253 & 636 & 1270 & 1910 & 2290 & 2430 \\
 HIP 51931 & 10 36 30.8 &  $-$13 50 36 & 948 & 1090 & 1340 & 1680 & 2090 & 2580 & 3060 \\
 TWA 7  & 10 42 30.1 &  $-$33 40 17 & 5 & 6 & 8 & 10 & 12 & 14 & 21 \\
 HIP 52462 & 10 43 28.3 &  $-$29 03 51 & 612 & 653 & 725 & 813 & 915 & 1040 & 1180 \\
 HIP 52787 & 10 47 31.2 &  $-$22 20 53 & 285 & 441 & 619 & 840 & 1120 & 1460 & 1860 \\
 HIP 53020 & 10 50 52.0 &  $+$06 48 29 & 4600 & 5160 & 6140 & 7500 & 8820 & 9550 & 9780 \\
 HIP 53486 & 10 56 30.8 &  $+$07 23 19 & 501 & 535 & 595 & 669 & 755 & 856 & 984 \\
 HD 95174 & 10 59 38.3 &  $+$25 26 15 & 20 & 23 & 358 & 3790 & 6890 & 8760 & 9380 \\
 HIP 53911  & 11 01 51.9 &  $-$34 42 17 & 5 & 6 & 8 & 10 & 12 & 14 & 20 \\
 HIP 54155 & 11 04 41.5 &  $-$04 13 16 & 24 & 49 & 125 & 251 & 392 & 576 & 870 \\
 HIP 54745 & 11 12 32.4 &  $+$35 48 51 & 412 & 440 & 491 & 553 & 624 & 709 & 820 \\
 TWA 14  & 11 13 26.2 &  $-$45 23 43 & 5 & 6 & 8 & 10 & 11 & 13 & 16 \\
 TYC 3825-716-1 & 11 20 50.5 &  $+$54 10 09 & 28 & 57 & 145 & 292 & 471 & 777 & 1320 \\
 TWA 13N  & 11 21 17.2 &  $-$34 46 46 & 5 & 6 & 8 & 10 & 12 & 14 & 21 \\
 TWA 8A  & 11 32 41.3 &  $-$26 51 56 & 5 & 6 & 8 & 10 & 12 & 14 & 21 \\
 HIP 57370 & 11 45 42.3 &  $+$02 49 17 & 741 & 790 & 876 & 983 & 1110 & 1260 & 1430 \\
 HIP 57494 & 11 47 03.8 &  $-$11 49 27 & 3870 & 4310 & 5040 & 5940 & 7040 & 8290 & 9050 \\
 TWA 9B  & 11 48 23.7 &  $-$37 28 49 & 5 & 6 & 8 & 10 & 12 & 14 & 21 \\
 HIP 57589  & 11 48 24.2 &  $-$37 28 49 & 5 & 6 & 8 & 10 & 12 & 14 & 21 \\
 HIP 57632 & 11 49 03.6 &  $+$14 34 19 & 30 & 33 & 37 & 42 & 50 & 1090 & 1510 \\
 HIP 58876 & 12 04 33.7 &  $+$66 20 12 & 19 & 39 & 97 & 194 & 304 & 437 & 635 \\
 G 123-7 & 12 09 37.3 &  $+$40 15 07 & 1530 & 1720 & 2060 & 2500 & 3020 & 3640 & 4170 \\
 TYC 4943-192-1 & 12 15 18.4 &  $-$02 37 28 & 109 & 115 & 123 & 131 & 141 & 304 & 484 \\
 HIP 59774 & 12 15 25.6 &  $+$57 01 57 & 411 & 433 & 465 & 501 & 537 & 572 & 601 \\
 TWA 25  & 12 15 30.7 &  $-$39 48 43 & 5 & 6 & 8 & 10 & 12 & 14 & 21 \\
 HIP 60074 & 12 19 06.5 &  $+$16 32 54 & 349 & 373 & 417 & 470 & 532 & 606 & 704 \\
 HIP 60661 & 12 25 58.6 &  $+$08 03 44 & 2880 & 3210 & 3780 & 4500 & 5330 & 6240 & 6900 \\
 HD 108767B & 12 29 50.9 &  $-$16 31 15 & 77 & 158 & 401 & 761 & 1240 & 1860 & 2520 \\
 TWA 20  & 12 31 38.1 &  $-$45 58 59 & 5 & 6 & 8 & 10 & 11 & 13 & 16 \\
 HIP 61174 & 12 32 04.2 &  $-$16 11 46 & 220 & 441 & 1100 & 2200 & 3310 & 3970 & 4190 \\
 TWA 10  & 12 35 04.3 &  $-$41 36 39 & 5 & 6 & 8 & 10 & 12 & 14 & 21 \\
 TWA 11B  & 12 36 00.6 &  $-$39 52 16 & 5 & 6 & 8 & 10 & 12 & 14 & 17 \\
 HIP 61498 & 12 36 01.0 &  $-$39 52 10 & 5 & 6 & 8 & 10 & 12 & 14 & 17 \\
 TWA 11A  & 12 36 01.0 &  $-$39 52 10 & 5 & 6 & 8 & 10 & 12 & 14 & 17 \\
 HIP 61960 & 12 41 53.1 &  $+$10 14 08 & 64 & 129 & 323 & 647 & 969 & 1160 & 1240 \\
 BD$+$60 1417 & 12 43 33.3 &  $+$60 00 53 & 35 & 72 & 184 & 376 & 694 & 1180 & 1780 \\
 HIP 62523 & 12 48 47.0 &  $+$24 50 25 & 1210 & 1290 & 1430 & 1590 & 1780 & 2000 & 2230 \\
 HIP 63076 & 12 55 28.5 &  $+$65 26 19 & 173 & 347 & 871 & 1740 & 2610 & 3140 & 3320 \\
 HIP 63317 & 12 58 32.0 &  $+$38 16 44 & 20 & 42 & 107 & 216 & 330 & 442 & 638 \\
 HIP 63584 & 13 01 46.9 &  $+$63 36 37 & 253 & 510 & 1280 & 2560 & 3850 & 4620 & 4880 \\
 HIP 63742 & 13 03 49.7 &  $-$05 09 43 & 113 & 117 & 123 & 130 & 138 & 146 & 164 \\
 HIP 64792 & 13 16 46.5 &  $+$09 25 27 & 166 & 180 & 205 & 236 & 268 & 303 & 352 \\
 HIP 65016 & 13 19 40.1 &  $+$33 20 48 & 3680 & 4080 & 4770 & 5620 & 6630 & 7810 & 8650 \\
 HIP 65515 & 13 25 45.5 &  $+$56 58 14 & 110 & 121 & 142 & 171 & 199 & 227 & 266 \\
 FH CVn & 13 27 12.1 &  $+$45 58 26 & 41 & 60 & 111 & 130 & 154 & 209 & 228 \\
 HIP 66252 & 13 34 43.2 &  $-$08 20 31 & 75 & 88 & 120 & 173 & 226 & 259 & 287 \\
 HIP 67092 & 13 45 05.3 &  $-$04 37 13 & 880 & 1380 & 2820 & 5210 & 7600 & 9040 & 9520 \\
 HIP 67412 & 13 48 58.2 &  $-$01 35 35 & 1280 & 1440 & 1740 & 2130 & 2600 & 3150 & 3660 \\
 HIP 69357 & 14 11 46.2 &  $-$12 36 42 & 1390 & 1570 & 1890 & 2310 & 2810 & 3400 & 3920 \\
 HD 234121 & 14 16 12.2 &  $+$51 22 35 & 16 & 34 & 87 & 177 & 279 & 413 & 623 \\
 HIP 69732 & 14 16 23.0 &  $+$46 05 18 & 67 & 136 & 343 & 689 & 1030 & 1240 & 1320 \\
 HIP 70952 & 14 30 46.1 &  $+$63 11 09 & 258 & 518 & 1300 & 2590 & 3890 & 4670 & 4930 \\
 HIP 71284 & 14 34 40.8 &  $+$29 44 42 & 211 & 427 & 1070 & 2140 & 3210 & 3850 & 4070 \\
 HIP 71395 & 14 36 00.6 &  $+$09 44 47 & 501 & 536 & 598 & 673 & 760 & 862 & 992 \\
 HIP 71631 & 14 39 00.2 &  $+$64 17 30 & 76 & 85 & 104 & 131 & 157 & 180 & 209 \\
 HIP 72146 & 14 45 24.2 &  $+$13 50 47 & 4270 & 4760 & 5580 & 6580 & 7770 & 8940 & 9470 \\
 HIP 72339 & 14 47 32.7 &  $-$00 16 53 & 2060 & 2320 & 2790 & 3380 & 4090 & 4880 & 5500 \\
 HIP 72567 & 14 50 15.8 &  $+$23 54 43 & 481 & 514 & 572 & 643 & 726 & 824 & 945 \\
 HIP 73996 & 15 07 18.1 &  $+$24 52 09 & 296 & 461 & 884 & 2360 & 3840 & 4720 & 5020 \\
 HIP 74045 & 15 07 56.3 &  $+$76 12 03 & 29 & 36 & 59 & 95 & 132 & 154 & 166 \\
 HIP 74702 & 15 15 59.2 &  $+$00 47 47 & 187 & 203 & 232 & 269 & 309 & 359 & 434 \\
 HIP 75829 & 15 29 23.6 &  $+$80 27 01 & 33 & 67 & 173 & 354 & 554 & 799 & 1130 \\
 HIP 76267 & 15 34 41.3 &  $+$26 42 53 & 407 & 430 & 464 & 500 & 535 & 569 & 593 \\
 HD 139084B  & 15 38 56.8 &  $-$57 42 19 & 17 & 18 & 19 & 21 & 22 & 24 & 29 \\
 HIP 76829 & 15 41 11.4 &  $-$44 39 40 & 35 & 72 & 182 & 368 & 564 & 716 & 789 \\
 HIP 77408 & 15 48 09.5 &  $+$01 34 18 & 769 & 819 & 908 & 1020 & 1150 & 1300 & 1470 \\
 HIP 77542 & 15 49 57.7 &  $-$03 55 16 & 4 & 4 & 4 & 5 & 5 & 5 & 5 \\
 HIP 78557 & 16 02 22.4 &  $+$03 39 07 & 11 & 24 & 62 & 128 & 210 & 434 & 3360 \\
 HIP 79762 & 16 16 45.3 &  $+$67 15 23 & 2510 & 2770 & 3230 & 3800 & 4470 & 5210 & 5780 \\
 HIP 79977 & 16 19 29.2 &  $-$21 24 13 & 6 & 7 & 8 & 10 & 11 & 13 & 16 \\
 HIP 80824 & 16 30 18.1 &  $-$12 39 45 & 3610 & 3980 & 4610 & 5370 & 6240 & 7190 & 7860 \\
 HIP 81084 & 16 33 41.6 &  $-$09 33 12 & 162 & 189 & 255 & 358 & 460 & 531 & 655 \\
 HIP 82587 & 16 52 58.1 &  $+$31 42 06 & 185 & 371 & 923 & 1840 & 2770 & 3320 & 3510 \\
 HIP 82688 & 16 54 08.1 &  $-$04 20 25 & 111 & 116 & 122 & 130 & 137 & 143 & 148 \\
 HIP 83494 & 17 03 53.6 &  $+$34 47 25 & 127 & 255 & 637 & 1270 & 1900 & 2280 & 2420 \\
 HD 155555C  & 17 17 31.3 &  $-$66 57 05 & 17 & 18 & 19 & 21 & 22 & 24 & 25 \\
 TYC 8728-2262-1  & 17 29 55.1 &  $-$54 15 49 & 18 & 19 & 20 & 24 & 82 & 118 & 130 \\
 HIP 86346 & 17 38 39.6 &  $+$61 14 16 & 111 & 116 & 122 & 129 & 137 & 143 & 148 \\
 HIP 87108 & 17 47 53.6 &  $+$02 42 26 & 53 & 108 & 274 & 550 & 825 & 993 & 1060 \\
 HIP 87322 & 17 50 34.0 &  $-$06 03 01 & 1130 & 1650 & 3070 & 5390 & 7690 & 9080 & 9540 \\
 HIP 87558 & 17 53 14.2 &  $+$06 06 05 & 268 & 538 & 1350 & 2710 & 4070 & 4880 & 5160 \\
 HIP 87579 & 17 53 29.9 &  $+$21 19 31 & 359 & 489 & 664 & 891 & 1170 & 1520 & 1920 \\
 HIP 87768 & 17 55 44.9 &  $+$18 30 01 & 29 & 59 & 151 & 305 & 484 & 706 & 1010 \\
 HD 164249B  & 18 03 04.1 &  $-$51 38 56 & 17 & 18 & 19 & 21 & 22 & 24 & 26 \\
 HIP 88848 & 18 08 16.0 &  $+$29 41 28 & 18 & 38 & 97 & 196 & 301 & 405 & 586 \\
 HIP 89005 & 18 09 55.5 &  $+$69 40 50 & 26 & 54 & 138 & 280 & 449 & 804 & 1360 \\
 HIP 91043 & 18 34 20.1 &  $+$18 41 24 & 30 & 36 & 52 & 78 & 104 & 120 & 133 \\
 HIP 92024  & 18 45 26.9 &  $-$64 52 17 & 17 & 18 & 19 & 21 & 22 & 24 & 27 \\
 TYC 9073-762-1  & 18 46 52.6 &  $-$62 10 36 & 17 & 18 & 20 & 21 & 24 & 253 & 331 \\
 TYC 7408-54-1  & 18 50 44.5 &  $-$31 47 47 & 17 & 18 & 19 & 21 & 23 & 88 & 153 \\
 HIP 92680  & 18 53 05.9 &  $-$50 10 50 & 17 & 18 & 19 & 21 & 22 & 23 & 25 \\
 HIP 92919 & 18 55 53.2 &  $+$23 33 24 & 55 & 63 & 85 & 121 & 156 & 179 & 200 \\
 HIP 93580 & 19 03 32.3 &  $+$01 49 08 & 113 & 117 & 124 & 132 & 144 & 1120 & 1680 \\
 HIP 95261  & 19 22 51.2 &  $-$54 25 26 & 18 & 19 & 20 & 24 & 451 & 722 & 815 \\
 HIP 95270  & 19 22 58.9 &  $-$54 32 17 & 17 & 18 & 19 & 21 & 22 & 24 & 27 \\
 HIP 95319 & 19 23 34.0 &  $+$33 13 19 & 5020 & 5660 & 6720 & 8040 & 9140 & 9690 & 9860 \\
 HIP 95793 & 19 29 01.0 &  $+$01 57 02 & 68 & 139 & 353 & 709 & 1070 & 1280 & 1360 \\
 HIP 97438 & 19 48 15.4 &  $+$59 25 22 & 14 & 29 & 74 & 152 & 242 & 355 & 532 \\
 2MASS J19560294-3207186  & 19 56 02.9 &  $-$32 07 19 & 17 & 18 & 19 & 21 & 23 & 301 & 511 \\
 TYC 7443-1102-1  & 19 56 04.4 &  $-$32 07 38 & 17 & 18 & 20 & 21 & 24 & 485 & 562 \\
 HIP 99273  & 20 09 05.2 &  $-$26 13 27 & 17 & 18 & 20 & 21 & 24 & 2760 & 4130 \\
 HIP 99711 & 20 13 59.8 &  $-$00 52 01 & 1670 & 1780 & 1960 & 2190 & 2440 & 2720 & 2990 \\
 HIP 101262 & 20 31 32.1 &  $+$33 46 33 & 696 & 898 & 1290 & 1810 & 2480 & 3270 & 3980 \\
 BD$+$05 4576 & 20 39 54.6 &  $+$06 20 12 & 47 & 95 & 124 & 141 & 334 & 560 & 854 \\
 HIP 102409 & 20 45 09.5 &  $-$31 20 27 & 17 & 18 & 19 & 21 & 22 & 24 & 29 \\
 HIP 104225 & 21 06 56.4 &  $+$69 40 29 & 4460 & 4980 & 5880 & 7060 & 8470 & 9420 & 9730 \\
 HIP 104308  & 21 07 51.2 &  $-$54 12 59 & 14 & 18 & 25 & 31 & 37 & 1420 & 2130 \\
 HD 201919 & 21 13 05.3 &  $-$17 29 13 & 113 & 117 & 123 & 130 & 137 & 145 & 152 \\
 HIP 105038 & 21 16 32.5 &  $+$09 23 38 & 1050 & 1190 & 1450 & 1790 & 2200 & 2690 & 3170 \\
 HIP 106231 & 21 31 01.7 &  $+$23 20 07 & 105 & 114 & 122 & 129 & 136 & 142 & 146 \\
 HIP 107345  & 21 44 30.1 &  $-$60 58 39 & 13 & 17 & 23 & 30 & 34 & 38 & 50 \\
 HIP 107350 & 21 44 31.3 &  $+$14 46 19 & 236 & 254 & 285 & 324 & 367 & 417 & 487 \\
 HIP 107649 & 21 48 15.8 &  $-$47 18 13 & 1680 & 1850 & 2170 & 2560 & 3030 & 3580 & 4020 \\
 HIP 108156 & 21 54 45.0 &  $+$32 19 43 & 3270 & 3620 & 4220 & 4960 & 5810 & 6720 & 7350 \\
 TYC 2211-1309-1 & 22 00 41.6 &  $+$27 15 14 & 14 & 18 & 20 & 23 & 85 & 166 & 211 \\
 V383 Lac & 22 20 07.0 &  $+$49 30 12 & 46 & 53 & 70 & 98 & 126 & 145 & 162 \\
 HIP 111449 & 22 34 41.6 &  $-$20 42 30 & 265 & 537 & 1350 & 2710 & 4060 & 4880 & 5150 \\
 TYC 9340-437-1  & 22 42 48.9 &  $-$71 42 21 & 17 & 18 & 19 & 21 & 22 & 24 & 29 \\
 HIP 112312  & 22 44 58.0 &  $-$33 15 02 & 17 & 18 & 19 & 21 & 22 & 24 & 27 \\
 TX PsA  & 22 45 00.1 &  $-$33 15 26 & 17 & 18 & 19 & 21 & 22 & 24 & 26 \\
 HIP 112909 & 22 51 53.5 &  $+$31 45 15 & 12 & 21 & 47 & 96 & 155 & 202 & 223 \\
 HIP 113020 & 22 53 16.7 &  $-$14 15 49 & 940 & 1440 & 2870 & 5250 & 7630 & 9050 & 9520 \\
 HIP 114066 & 23 06 04.8 &  $+$63 55 34 & 113 & 117 & 123 & 130 & 137 & 145 & 151 \\
 HIP 115147 & 23 19 26.6 &  $+$79 00 13 & 21 & 23 & 27 & 32 & 43 & 232 & 356 \\
 HIP 115162 & 23 19 39.6 &  $+$42 15 10 & 112 & 116 & 123 & 130 & 137 & 144 & 150 \\
 TYC 5832-666-1  & 23 32 30.9 &  $-$12 15 51 & 17 & 18 & 19 & 21 & 22 & 24 & 29 \\
 HIP 116215 & 23 32 49.4 &  $-$16 50 44 & 19 & 20 & 436 & 789 & 1080 & 1400 & 1710 \\
 HIP 116384 & 23 35 00.3 &  $+$01 36 19 & 29 & 59 & 147 & 289 & 459 & 756 & 1260 \\
 HIP 116805 & 23 40 24.5 &  $+$44 20 02 & 21 & 25 & 34 & 130 & 293 & 393 & 430 \\
 HIP 117410 & 23 48 25.7 &  $-$12 59 15 & 26 & 54 & 138 & 278 & 445 & 751 & 1260 \\
 HIP 118121  & 23 57 35.1 &  $-$64 17 54 & 13 & 17 & 23 & 30 & 34 & 38 & 44 
\enddata
\label{tab:allages}
\end{deluxetable*}

\section{Substellar Cooling Models} \label{sec:cooling_models}

Objects below $\sim$80 $M_{\rm Jup}$ never attain the central densities and temperatures necessary to stabilize themselves by hydrogen fusion; instead, they simply cool and fade below the detectability limits of even the largest telescopes.  Substellar cooling models simulate the structure and atmosphere of such an object as a function of its mass, age, and initial thermodynamic state, producing grids of spectra and luminosities.  Many such models have been developed recently and are widely used \citep[e.g.][]{Chabrier+Baraffe+Allard+etal_2000, Allard+Hauschildt+Alexander+etal_2001, Baraffe+Chabrier+Barman+etal_2003, Marley+Fortney+Hubickyj+etal_2007, Allard+Homeier+Freytag_2011, Spiegel+Burrows_2012}.  These models make a range of assumptions about dust settling, cloud structure, chemical composition, and initial thermodynamic state.  

The formation mechanism of planetary mass objects is still debated, and can strongly impact their brightness at young ages.  Direct gravitational collapse forms objects on a dynamical timescale \citep{Boss_2000}, giving them little time to radiate away their heat of formation.  In contrast, an object formed by core-accretion accretes its gas over a viscous time, and is likely to radiate away much of its initial energy in an accretion shock \citep{Hubickyj+Bodenheimer+Lissauer_2005, Marley+Fortney+Hubickyj+etal_2007}.  However, the identification of a core-accretion formation scenario with a ``cold start'' is far from certain; a subcritical shock, for example, could prevent efficient cooling \citep{Bromley+Kenyon_2011}, while a very massive core could provide a substantial reservoir of heat \citep{Mordasini_2013}.  Observational evidence also disfavors a very cold start for many imaged planets \citep{Marleau+Cumming_2014}.  Massive brown dwarfs are generally accepted to form by direct gravitational collapse, either in a disk or a molecular cloud \citep{Bate+Bonnell+Bromm_2002}.

In this work, we explore whether the entire range of long-period substellar objects is compatible with a single distribution function.  We therefore seek a single model applicable over this entire range, including the more massive brown dwarfs.  Such a model would imply that all of these objects form by direct gravitational collapse, and therefore, form hot.  We adopt the recent BT-Settl models \citep{Allard+Homeier+Freytag_2011}, which incorporate clouds and dust settling in the appropriate physical regimes, and are intended to be valid all the way from $\sim$1 M$_{\rm J}$ up to stellar masses.  BT-Settl reproduces the earlier COND/DUSTY hot-start models of \cite{Baraffe+Chabrier+Barman+etal_2003} and \cite{Chabrier+Baraffe+Allard+etal_2000} in the applicable limits.  

We also explore the effects of the formation model on our results restricted to the planetary mass regime, below the deuterium-burning limit of $\sim$13 M$_{\rm J}$.  In this case, we parametrize our ignorance using the ``warm-start'' models of \cite{Spiegel+Burrows_2012}, henceforth SB12.  These model grids include an additional parameter---initial entropy---at each age, mass, and atmospheric composition.  We restrict our analysis to a fiducial model with clouds and 3 times Solar metallicity.  

Under any formation scenario, the initial entropy is likely to be a function of mass.  More massive objects can attain a higher entropy while still permitting hydrostatic equilibrium, resulting in hotter starts.  However, they can also radiate more effectively (and for much longer) under a core-accretion scenario, producing an initially colder object.  We parametrize the range of initial thermodynamic states using a single ``warmth parameter,'' $\eta$, to fill the space between the high- and low-entropy envelopes given by SB12:
\begin{equation}
S_{\rm init}(M) = \eta S_{\rm init,\,max}(M) + (1 - \eta) S_{\rm init,\,min}(M)~,
\end{equation}
with $\eta \in [0,\,1]$.

Much uncertainty remains throughout the modeling process, and we make no claim that our use of the BT-Settl and SB12 models here exhausts parameter space.  We adopt the BT-Settl models in an attempt to include the entire range of substellar masses in a consistent manner, while we use the SB12 models to smoothly explore the effect of the initial thermodynamic conditions.  Substellar luminosities depend strongly on both mass and age, limiting (at least somewhat) the impact of systematic errors in predicted luminosity on the expected distribution of objects detected by high-contrast imaging.

\section{Statistical Framework} \label{sec:framework}

High-contrast imaging data provide a detection limit (or a detected companion) everywhere on the field-of-view; we want to compare these data to predictions made by models of exoplanet formation and evolution.  The probability of a planet existing at any individual location around a particular star is tiny.  Hence, in the absence of any other information about the star, the planet's existence is a Poisson process.  If we denote by $\lambda_i$ the expected number of substellar companions that exist and are detectable at a given position and luminosity $i$, given some detection limits and model of planet properties, then the probability of such a companion existing is $\lambda_i e^{-\lambda_i}$ ($\approx \lambda_i$), while the probability of its not existing (either because there is no companion or because it is too faint) is $e^{-\lambda_i}$ ($\approx 1 - \lambda_i$).  Here, $i$ runs over both position and substellar luminosity, and the $\lambda_i$ are functions of the substellar companion distribution function, cooling models, and survey contrast.  

The approximation of planet occurrence as an independent Poisson process at every point around every star surveyed breaks down with multiple systems or when other information, such as stellar binarity, makes some positions more dynamically favorable than others.  The famous four (or more) planet system HR~8799 illustrates this empirically.  Various studies have found that few stars (typically no more than 5--10\%) host massive ($\gtrsim$5 $M_{\rm Jup}$), long-period exoplanets \citep[e.g.][]{Lafreniere+Doyon+Marois+etal_2007, Vigan+Patience+Marois+etal_2012, Nielsen+Liu+Wahhaj+etal_2013, Chauvin+Vigan+Bonnefoy+etal_2014}.  If planet occurrence is independent at different locations, the number of single planet systems should exceed the number of four-planet systems by at least a factor of $10^3$--$10^4$, while the number of two-planet systems should be larger than HR8799-like systems by a factor of at least $10^2$--$10^3$.  The fact that we have not found a single system with two $\sim$5--10 $M_{\rm Jup}$ companions argues powerfully against independent occurrence probabilities, even before considering theoretical arguments.  

As high-contrast surveys improve in sensitivity and target hundreds of new stars, the number of known multiple systems will almost certainly rise.  However, the data sets we consider here lack any detected multiple planet systems, freeing us (for now) from the statistical and theoretical problems they present.  With the caveat that our statistical framework must be abandoned or at least modified if systems like HR 8799 are to be included, we therefore proceed by assuming planet occurrence to be an independent Poisson process.  The likelihood function ${\cal L}$, the probability density of the given data set (detected companions plus sensitivity limits) given a model of planet frequency and properties, is then the product of the planet-occurrence probabilities over all elements of the fields-of-view of all stars, and over all substellar luminosities:
\begin{equation}
{\cal L} = \prod_{i} e^{-\lambda_i} \times \prod_{{\rm obj}~j} \lambda_j e^{-\lambda_j}~.
\end{equation}
The index $i$ runs over all locations around all stars and over all planet luminosities, apart from those positions and luminosities at which a companion was seen.  The index $j$ only runs over the pixels and luminosities where substellar companions were discovered.  
The latter term is the likelihood of finding all of the detected companions at their actual positions with their observed luminosities.  The exponential term in the product over $j$ combines with the identical term in the product over $i$ to give a product over all pixels around all stars and over all planet luminosities, {\it including those where companions were discovered.}  The likelihood is more conveniently written logarithmically, and reduces to
\begin{equation}
\ln {\cal L} = -\sum_i \lambda_i + \sum_{{\rm obj}~j} \ln \lambda_j
= -\langle N_{\rm obj} \rangle + \sum_{{\rm obj}~j} \ln \lambda_j~,
\label{eq:log_like1}
\end{equation}
where now $i$ runs over all pixels around all stars and over all companion luminosities.  The first term, $\lambda_i$ summed over all planet positions and luminosities, is simply the expected number of detectable companions, $\langle N_{\rm obj} \rangle$, given the substellar distribution function.  

The substellar distribution function, from which the $\lambda_i$ and $\lambda_j$ are computed, can depend arbitrarily on parameters like the host stellar type, companion mass, separation, host metallicity, etc., as we discuss in the next section.  Regardless of its functional form, the distribution function may be multiplied by a constant $A$, which multiplies all probabilities $\lambda_i$ and $\lambda_j$ (and, by extension, $\langle N_{\rm obj}\rangle$) by the same factor.   This normalization constant is generally not known {\it \`a priori} and must be fit by our analysis; it gives, e.g., the total number of companions per star.  Such a constant enters the likelihood calculation, Equation \eqref{eq:log_like1}, in a particularly simple way.  Multiplying the distribution function, and by extension, all $\lambda_i$ and $\lambda_j$, by a constant $A$, the likelihood function becomes
\begin{equation}
\ln {\cal L} = -A \langle N_{\rm obj} \rangle + N_{\rm obs} \ln A + \ln \lambda_j~,
\label{eq:gammafunc}
\end{equation}
where $N_{\rm obs}$ is the number of companions actually observed.  If we hold all other parameters of the distribution fixed while determining the normalization, $\lambda_j$, $\langle N_{\rm obj}\rangle$, and $N_{\rm obs}$ are all constant, and Equation \eqref{eq:gammafunc} is a gamma distribution.  The maximum likelihood value of $A$ is, as expected, $N_{\rm obs} / \langle N_{\rm obj} \rangle$.   In other words, the maximum likelihood value of $A$ is the one that makes the predicted number of detections $\langle N_{\rm obj}$ equal to $N_{\rm obs}$, the actual number of substellar companions seen in the survey.

Setting $A$ equal to its maximum likelihood value, the likelihood function now reads
\begin{equation}
{\cal L} = \exp \left(-N_{\rm obs} \right) \prod_{{\rm obj}~j} \frac{\lambda_j}{\langle N_{\rm obj} \rangle}~.
\label{eq:log_like2}
\end{equation}
As a result, all we need to calculate from the substellar distribution function, cooling model, and sensitivity curves are the probabilities of detecting companions at their observed positions with their observed properties, and the total number of expected detections.  In the next section, we discuss the computation of $\lambda_j$ and $\langle N_{\rm obj} \rangle$ in the case of a power-law companion distribution function.

\subsection{The Exoplanet Distribution Function} \label{subsec:dist_func}

Neglecting multiplicity, the exoplanet distribution function extends over three parameters of interest: mass $M$, semimajor axis $a$, and eccentricity $e$.  Other orbital or orientation parameters are randomly distributed, with those distributions determined by geometry (assuming the systems to be randomly oriented).  The distribution function is usually assumed to be separable, i.e.,
\begin{equation}
p(M, a, e) = p(M) p(a) p(e)~.
\end{equation}
While this is unlikely to be true in reality, and may not even be a particularly good approximation, we adopt it for lack of a specific and well-motivated alternative.  

When we observe a companion, the only measurements we can usually make are of its projected separation and its luminosity at various wavelengths.  It is therefore the distribution of projected separation, a function of both the semimajor axis and eccentricity distributions, that is relevant to imaging surveys.  The choice of eccentricity distribution turns out to be of secondary importance to this distribution.  Indeed, in the special case that the semimajor axis is a power law truncated well outside the separations being probed, the eccentricity distribution drops out altogether--$p(M, a)$ is independent of $p(e)$ (see the Appendix for details).

We derive the distribution of projected separation based on an eccentricity distribution uniform in the range from 0 to 0.8 \citep{Cumming+Butler+Marcy+etal_2008}; the result is nearly identical to that obtained using a Rayleigh distribution in eccentricity with $\sigma = 0.3$ \citep{Juric+Tremaine_2008}.  For completeness, we also include a distribution based on $p(e)=2e$, the theoretical distribution expected if the phase space density of companions is a function of energy only \citep{Ambartsumian_1937}.  Full details and piecewise analytic fitting functions are provided in the Appendix.

We take the exoplanet mass distribution to be a power law $p(M) \propto M^\beta$, and use cooling models of substellar objects to convert mass and age to a luminosity $L$ in a given bandpass.  The azimuthal variance in the detection limit is dominated by subtraction residuals from the stellar PSF and by read noise---photon noise from the companion itself is nearly always negligible.  These noise distributions are typically very nearly Gaussian.  With a detection threshold of $N_{\sigma}$ (usually 5--6), the probability of detecting a companion is therefore 
\begin{align}
p({\rm detect}|L) &= p(L + x > N_\sigma \cdot \sigma) \nonumber \\
&= \frac{1}{2} + \frac{1}{2} {\rm erf} \left(\frac{(L/L_{\rm lim}-1) N_\sigma}{\sqrt{2}} \right)~,
\label{eq:erf}
\end{align}
where $x$, a Gaussian random variable with variance $\sigma^2$, represents the noise, $L_{\rm lim}$ is the formal $N_\sigma \sigma$ detection threshold, and $\sigma^2 = \sigma^2(D)$ is the annular variance in the final ADI-processed image.  The total differential number of sources expected at angular separation $D$ is 
\begin{equation}
dN(D) = p(D) \, dD  \int_0^\infty \frac{dN}{dM} \frac{dM}{dL} p({\rm detect}|L)\,dL~.
\label{eq:dp_dD}
\end{equation}
The substellar-cooling models consist of grids in age and mass, which we interpolate using piecewise power laws.  We, therefore, evaluate Equation \eqref{eq:dp_dD} for the special case of a power law $dM/dL$; it reduces to a function of a single parameter with variable limits of integration.  We tabulate these values, enabling us to evaluate Equation \eqref{eq:dp_dD} to a typical accuracy of much better than $0.1$\% with no more than a few array lookups, $\lesssim$20 floating point operations, and at most one call to a special function like the error function.  The full derivation is given in the Appendix.  

Finally, we integrate Equation \eqref{eq:dp_dD} over projected separations $D$ to obtain the total number of expected detections around an object, giving us $\langle N_{\rm obj} \rangle$ in Equation \eqref{eq:log_like2}.  We perform the integral using the trapezoidal rule in logarithmic separation, interpolating from the input contrast curves using cubic splines.  Our use of cubic splines virtually guarantees the lack of a finite second derivative, irrespective of any intrinsic features of the contrast curves themselves, and makes the trapezoidal rule first order.  We therefore use Richardson extrapolation with a fit linear in step size, and accept the result of the extrapolation when it agrees with the previous extrapolated result (using half as many function evaluations) within a factor of $3 \times 10^{-3}$.  This tolerance allows relatively few function evaluations, while preventing the integration from dominating our error budget.  It generally requires using $\sim$30--100 points in angular separation, at a cost of $\sim$0.05--0.1 ms per star on a single 3.3 GHz thread.  Processing an entire large survey of 200 stars with a single set of test parameters thus requires about 10--20 ms, an enormous improvement over Monte Carlo methods that generate millions of planets around each star.  

After evaluating $\langle N_{\rm obj} \rangle$ by integrating Equation \eqref{eq:dp_dD} and summing over stars, we need to evaluate the probabilities $\lambda_j$ of detectable companions existing at their observed positions $D_j$ and luminosities $L_j$.  Substellar companions are inevitably followed up many times, making the uncertainties on their luminosities very small.  The photometric uncertainties are often dominated by scatter in the host star's photometry and variation in the AO performance across the field, bearing little relation to the annular standard deviation in the discovery image.  The probabilities $\lambda_j$, taking the $L_j$ as known, become
\begin{equation}
\lambda_j = p({\rm detect}|D_j,L_j) \left( \frac{dN}{d\ln L\,dD} d\ln L\,dD \right) \bigg|_{D_j,L_j} ~.
\label{eq:like_exoplanet}
\end{equation}
This is equivalent to using Monte Carlo to find the number of companions in a small interval $\Delta {\rm mag}$.  The detection probability is a function only of $L_j$ (which is observed).  Because it is independent of the substellar distribution function, the detection probability contributes to an overall scaling of the likelihood function and has no effect on our analysis.  

In practice, it is problematic to use Equation \eqref{eq:like_exoplanet} as written because the derivative can be discontinuous or even, near the deuterium-burning limit, singular.  We therefore integrate it over a small interval in $\ln L$:
\begin{equation}
\lambda_j \propto \int_{x_1}^{x_2} \frac{dN}{d\ln L\,dD} d\ln L~,
\label{eq:like_exoplanet2}
\end{equation}
We adopt 0.1 magnitudes ($\sim$10\% in luminosity) as our fiducial half-width $(x_2 - x_1)/2$, which is comparable to the photometric errors on well-characterized companions like GJ 504b \citep{Kuzuhara+Tamura+Kudo+etal_2013} and $\beta$ Pictoris b \citep{Bonnefoy+Boccaletti+Lagrange+etal_2013}.  Though this width is somewhat arbitrary, we note that half-widths from 0.01 to 0.2 magnitudes give indistinguishable best-fit parameters and confidence intervals in our final analysis.  Constant factors, like a variable increment in $dD$, do not affect our result; they merely add a constant to the log likelihood (equivalently, they multiply the likelihood by a constant).

All of the approximations and tabulations described above introduce errors of $\lesssim$1\% (and substantially less for most parameter values of interest), which are smaller than those introduced by interpolating substellar cooling models in mass and age.  The tabulated cooling models (Section \ref{sec:cooling_models}) and stellar ages (Section \ref{sec:bayes_ages}) dominate the errors in our analysis.

\subsection{Model Constraints} \label{subsec:model_constraints}

Our model for the exoplanet distribution function, 
\begin{equation}
N(M, a) = A M^\beta a^\alpha,
\end{equation}
with 
\begin{equation}
a_{\rm min} < a < a_{\rm max} \quad {\rm and} \quad M_{\rm min} < M < M_{\rm max},
\end{equation}
has seven free parameters, not counting the substellar cooling model:
\begin{enumerate}
\item The normalization of the distribution function, or, equivalently, the fraction of stars hosting substellar companions with a given range of properties;
\item $\alpha$, the slope of the semimajor axis distribution function;
\item $\beta$, the slope of the mass distribution function;
\item $M_{\rm min}$, the minimum mass of the population;
\item $M_{\rm max}$, the maximum mass of the population;
\item $a_{\rm min}$, the inner edge of the semimajor axis distribution; and
\item $a_{\rm max}$, the outer edge of the semimajor axis distribution. 
\end{enumerate}
The normalization could also be a function of the host stellar mass, while $a_{\rm min}$ and $a_{\rm max}$ are both likely to scale with stellar mass (as might $M_{\rm min}$ and $M_{\rm max}$), which could add even more parameters.  

Unfortunately, our sample contains far too few detections---$\kappa$ And b, plus two brown dwarf companions in the Pleiades and two companions to NICI targets---to provide meaningful constraints on the full companion distribution function.  Instead, we ask two much simpler questions:
\begin{itemize}
\item Out to what semimajor axis could the exoplanet distribution function measured by RV surveys extend?
\item Can we use a single power law to describe the full population of objects from $\sim$5 M$_{\rm J}$ to massive brown dwarfs?
\end{itemize}
In this way, we attempt to address the question of whether the planets now being discovered by direct imaging represent a new population, distinct from both an extrapolation of the RV sample and of the more massive brown dwarfs.  

Even addressing these more limited questions poses considerable difficulties.  The distribution function of RV planets around solar-type FGK stars has been reasonably well-measured \citep{Cumming+Butler+Marcy+etal_2008}; however, it depends both on stellar mass and on metallicity \citep{Johnson+Butler+Marcy+etal_2007, Fischer+Valenti_2005}.  
We lack metallicity data on many of our stars, and the sample, being largely young, spans a relatively narrow range in metallicity.  We therefore make no attempt to take stellar enrichment into account.  We do, however, attempt to at least qualitatively correct for host stellar mass, by assuming the planet-hosting probability to be directly proportional to the stellar mass in units of $M_\odot$ (a crude fit to the histogram presented in \citealt{Johnson+Butler+Marcy+etal_2007}).  This fits with the observed correlation of disk mass with stellar mass \citep{Andrews+Rosenfeld+Kraus+etal_2013}, though a correlation between stellar and companion {\it mass} may be a more natural choice.  The latter is equivalent to using an upper limit of the mass distribution that depends on host stellar mass, and emphasizes the artificial nature of using a power law truncated at the deuterium burning limit.  

The substellar cooling model appropriate to the RV population is hotly debated (see the discussion in Section \ref{sec:cooling_models}).  Objects formed by direct gravitational collapse do not have time to radiate away their heat of formation, while objects formed by core-accretion could lose much of their initial entropy in an accretion shock.  We parametrize our ignorance using the SB12 warm-start models, using a variable warmth parameter $\eta \in [0, 1]$ to interpolate between the warmest and coldest starts for each mass.  We also include the BT-Settl hot-start models \citep{Allard+Homeier+Freytag_2011}, which apply over the entire range of masses from $\sim$1 M$_{\rm J}$ to the hydrogen burning limit.

For our fits to a brown-dwarf-like distribution, we set the limits on the semimajor axis distribution to be 1--1000 AU, well outside the regions typically probed around our targets, and set the lower mass cutoff to be 5 M$_{\rm J}$, consistent with the minimum mass predicted by models of disk instability at wide separations \citep{Rafikov_2005} and fragmentation in molecular clouds \citep{Low+Lynden-Bell_1976, Bate_2003}.  Unfortunately, the upper mass limit of this distribution is uncertain: the upper mass cutoff for disk fragmentation could differ from that for cloud fragmentation.  However, our data lack the power to determine this cutoff, which we provisionally set at 70 M$_{\rm J}$.  A brown-dwarf like distribution is generally accepted to form by direct gravitational collapse.  The hot-start BT-Settl models thus provide an appropriate benchmark to constrain the properties of this distribution.  

Finally, we adopt uniform priors on $\alpha$ and $\beta$ and reinterpret the likelihood function as an unnormalized posterior probability distribution.  Using Monte Carlo to compute the normalization, we may then place constraints on $\alpha$ and $\beta$.

\section{Results and Discussion} \label{sec:results}

\begin{figure}
\includegraphics[width=\linewidth]{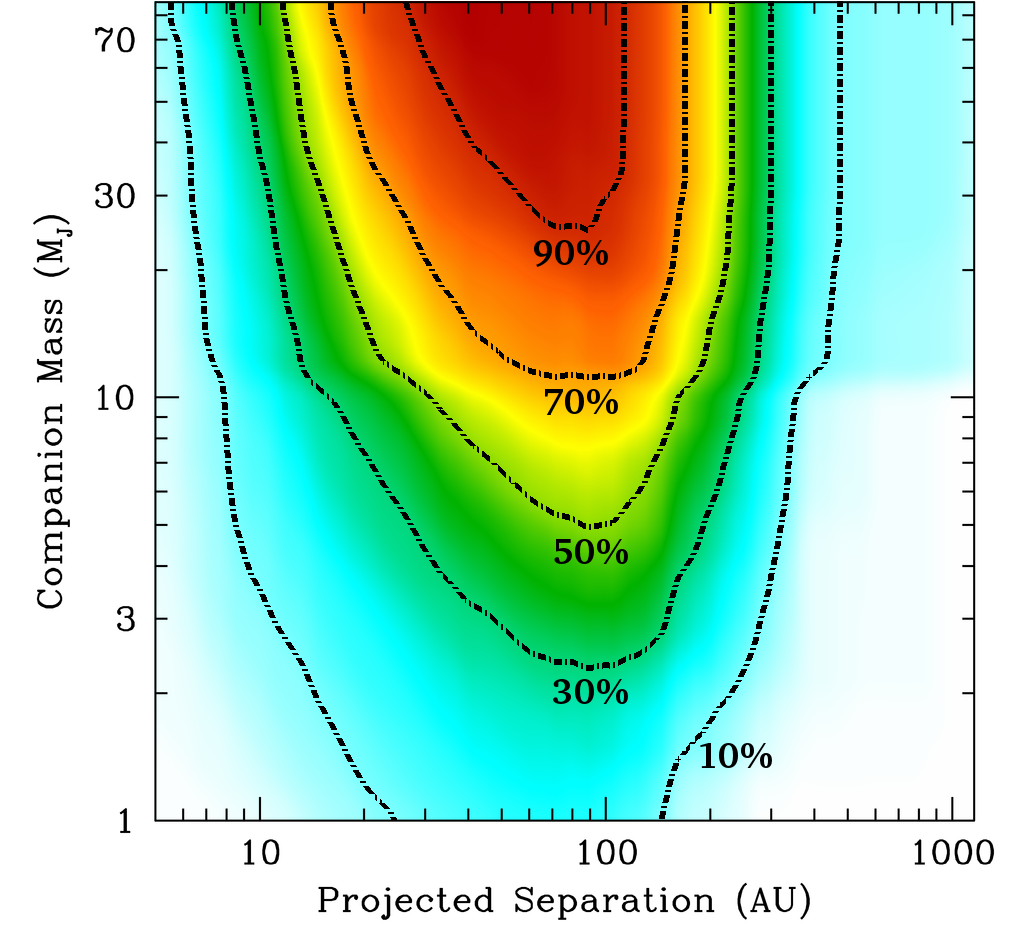}
\includegraphics[width=\linewidth]{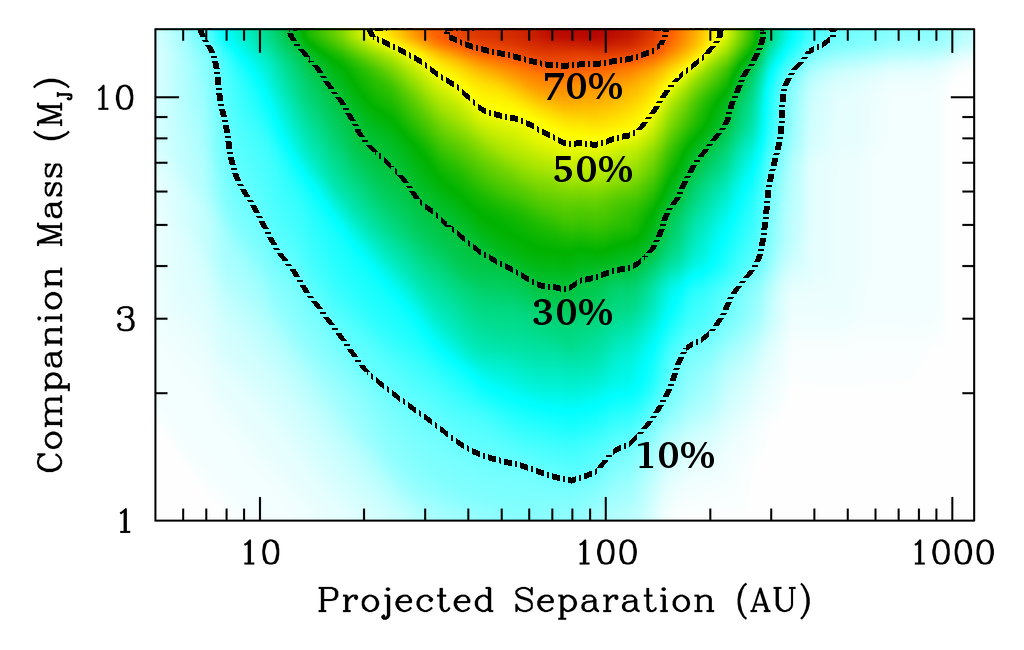}
\caption{The completeness of our combined survey, calculated using the BT-Settl hot-start models (top panel) and a warm start model (initial entropy halfway between the hottest and coldest starts), and weighting all stars equally.  At separations of 50--100 AU, we are about 70\% complete at the deuterium burning threshold of $\sim$13 M$_{\rm J}$ under both models.  }
\label{fig:completeness_bt-settl}
\end{figure}

Figure \ref{fig:completeness_bt-settl} shows the completeness of our data set given our age estimates and assuming either the hot-start BT-Settl cooling models \citep{Allard+Homeier+Freytag_2011}, or a warm-start model (SB12) with an initial entropy halfway between the coldest and hottest starts ($\eta=0.5$).  At separations of 50--100 AU, we are sensitive to $\sim$60\% of objects near the deuterium burning limit.  This value is higher around stars closer to Earth, which tend to be of lower mass, and around stars reliably identified with a young MG.

In this section, we use the detection limits to determine the limits we can place on an extrapolation of the distribution function measured by radial velocities.  We then proceed to model the entire substellar distribution function using a single power law, and determine its parameters and their confidence intervals. 

\subsection{Limits on an RV Distribution Function} \label{subsec:rv_limits}

The distribution function as measured by \cite{Cumming+Butler+Marcy+etal_2008} follows $M^{-1.3} a^{-0.6}$ from 0.5 to 10 M$_{\rm J}$ and from 0.03 to 3 AU.  We extrapolate this up to the deuterium burning limit of $\sim$13 M$_{\rm J}$ and out to a semimajor axis of $a_{\rm max}$, which we seek to constrain.  The normalization of the \citeauthor{Cumming+Butler+Marcy+etal_2008} distribution function is not free: 10.5\% of 1 $M_\odot$ stars have a planet with a mass between 0.3 and 10 $M_{\rm Jup}$ and a semimajor axis between 0.03 and 3 AU.  With the normalization fixed, we integrate Equation \eqref{eq:dp_dD} over separations and sum over stars to get the expected number of detections, $\langle N_{\rm obj} \rangle$.  We then compare $\langle N_{\rm obj} \rangle$ to the actual number of planet candidates in our sample using the Poisson distribution.

If an extrapolated RV planet distribution is to explain recent discoveries like GJ 504b, HR 8799b, $\kappa$ And b, and HD 95086b, $a_{\max}$ must be $\gtrsim$50 AU, where these companions have been found.  The sample we present here, depending on whether $\kappa$ And b is hypothesized to arise from an RV-like distribution, has at most one detection in total, and zero around FGKM stars.  If $\kappa$ And b is more than $\sim$15 M$_{\rm J}$, as would be the case if, as suggested by \cite{Hinkley+Pueyo+Faherty+etal_2013} and \cite{Bonnefoy+Currie+Marleau+etal_2014}, it is older than the $\sim$30 Myr implied by membership in the Columba MG \citep{Carson+Thalmann+Janson+etal_2013}, we can exclude models that predict more than 3 detections with 95\% confidence.  If $\kappa$ And b is considered to be a candidate member of an extrapolated RV distribution, the 2$\sigma$ threshold rises to 4.7 predicted detections.  We truncate the distribution at the deuterium burning threshold of 13 M$_{\rm J}$ for comparison to previous results and to facilitate the use of the SB12 models, which are only calculated for masses up to 15 M$_{\rm J}$.

\begin{figure}
\includegraphics[width=\linewidth]{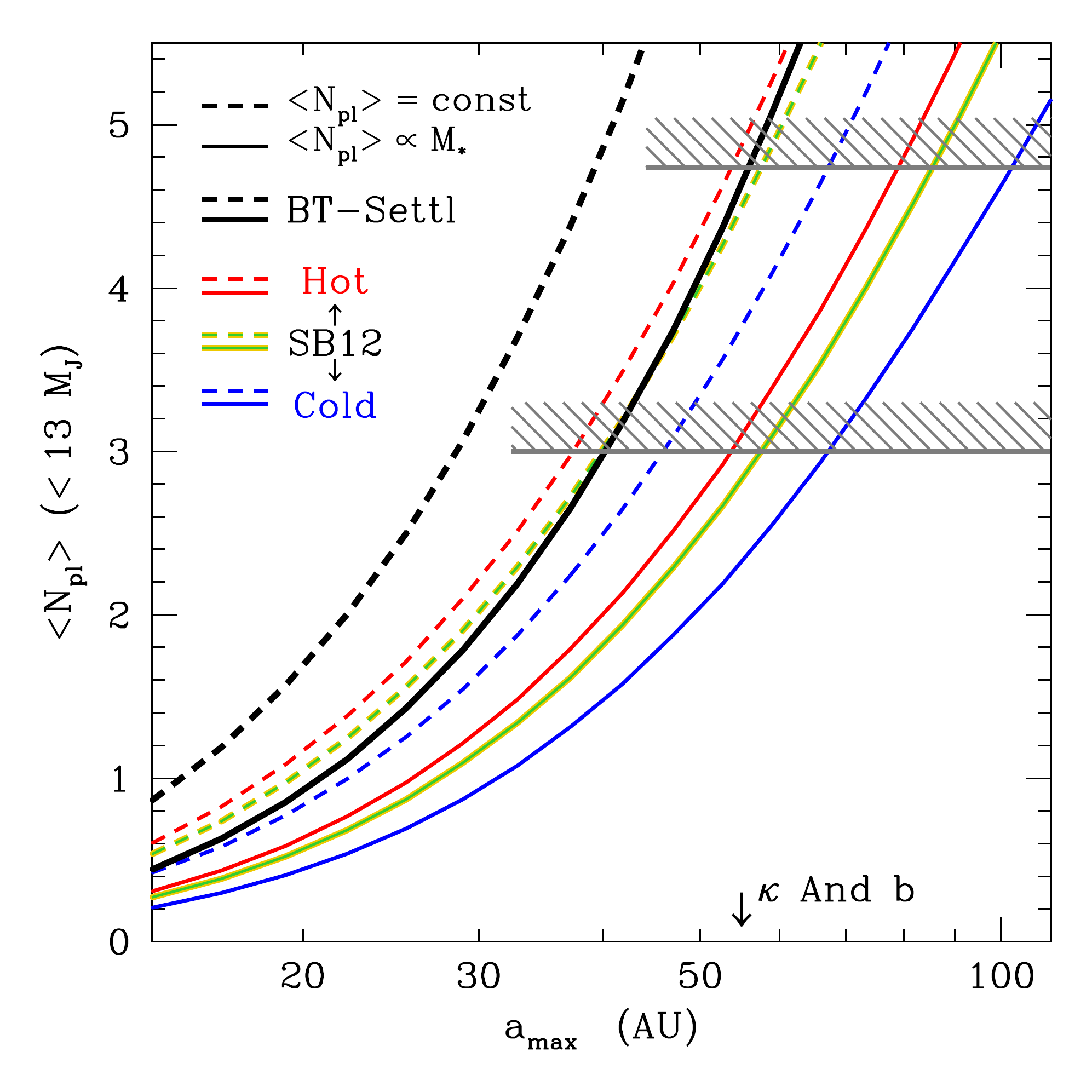}
\caption{Expected number of planetary-mass detections as a function of semimajor axis cutoff and cooling model, computed by extrapolating the measured distribution function for RV planets \citep{Cumming+Butler+Marcy+etal_2008}.  The solid lines assume the number of companions to be proportional to host stellar mass, while the dashed lines assume no proportionality.  The gray hatched regions are excluded at 2$\sigma$ assuming $\kappa$ And b to be drawn from this distribution (top, $\langle N_{\rm pl} \rangle \approx 4.7$), or to belong to a separate distribution (bottom, $\langle N_{\rm pl} \rangle \approx 3$)}
\label{fig:a_limits}
\end{figure}

Figure \ref{fig:a_limits} shows the predicted number of detections as a function of $a_{\rm max}$ for the BT-Settl model and for SB12 warm-start models spanning the range from hot to cold starts.  For the BT-Settl hot start models, the 2$\sigma$ upper limit on $a_{\rm max}$ varies from 30 to 50 AU, depending on whether we scale the companion frequency with stellar mass and whether we consider $\kappa$ And b to arise from this distribution.  If we adopt the SB12 models, the equivalent 2$\sigma$ upper limits vary from 40 to 70 AU as long as we do not consider the very coldest start models (those with $\eta \lesssim 0.1$).  If we do adopt these cold-start models, the upper limit on $a_{\rm max}$ rises to as much as $\sim$150 AU.  We note that cold-start models would predict a mass for $\kappa$ And b well in excess of the deuterium burning limit \citep{Carson+Thalmann+Janson+etal_2013} and would justify that object's classification as a nonmember of this distribution, regardless of its membership in Columba.

Our finding that the RV distribution function of \cite{Cumming+Butler+Marcy+etal_2008} cannot be extrapolated past a semimajor axis of $\sim$30--70 AU for most assumptions about the substellar cooling model is similar to the earlier results of \cite{Nielsen+Close_2010} and \cite{Chauvin+Lagrange+Bonavita+etal_2010}.  \cite{Nielsen+Close_2010} used a smaller sample of 118 targets dominated by the GDPS, finding limits on $a_{\rm max}$ from $\sim$65--200 AU depending on the substellar cooling model and on the correlation between planet frequency and stellar mass.  \cite{Chauvin+Lagrange+Bonavita+etal_2010} used 88 young stars to constrain this outer limit to be $\sim$80 AU, again depending on the cooling model and the details of the distribution function.  These results are in mild tension with the discovery of objects like HR 8799b, GJ 504b. and HD 95086b. which all lie at separations of $\gtrsim$50 AU.  There may be even more tension with the form of the mass distribution, which in RV surveys, is a power law increasing sharply towards low masses.  \cite{Wahhaj+Liu+Nielsen+etal_2013} found a positive power law index for the distribution in planet mass (more massive objects are more common), in sharp disagreement from the RV findings.  This result, however, was driven by their inclusion of the four HR 8799 planets as independent detections and their lack of any detected companions $\lesssim$5 M$_{\rm J}$.  Excluding HR 8799 from the \cite{Wahhaj+Liu+Nielsen+etal_2013} analysis weakens this finding considerably.

\subsection{A Single Substellar Distribution Function} \label{subsec:onedist}

The preceding analysis artificially separates objects below and above the deuterium burning threshold, making the classification of substellar companions like $\kappa$ And b problematic.  It also does not consider the properties of the detected companions.  
We now consider a distribution function extending across the deuterium-burning threshold, up to the hydrogen burning limit of $\sim$70 $M_{\rm Jup}$.  The predicted probability density of detections may be projected onto substellar mass and semimajor axis, and compared with our sample.  For this exercise, and for the statistical analysis that follows, we add two additional substellar companions discovered by HiCIAO: GJ 758B \citep{Thalmann+Carson+Janson+etal_2009, Janson+Carson+Thalmann+etal_2011}, a $\sim$30 M$_{\rm J}$ brown dwarf around an old G star first imaged during HiCIAO commissioning, and GJ 504b \citep{Kuzuhara+Tamura+Kudo+etal_2013, Janson+Brandt+Kuzuhara+etal_2013}, a $\sim$3--8 M$_{\rm J}$ companion to an active field G star discovered during the full SEEDS survey.  By doing this, we assume that the contrasts, distances, and ages of the as-yet-unpublished HiCIAO data are similar to those presented in Section \ref{sec:data}.  In reality, the unpublished stars stars represent a combination of very young members of starforming regions and nearby stars with a wide range of ages, a heterogeneity to that of our combined sample.

\begin{figure}
\includegraphics[width=\linewidth]{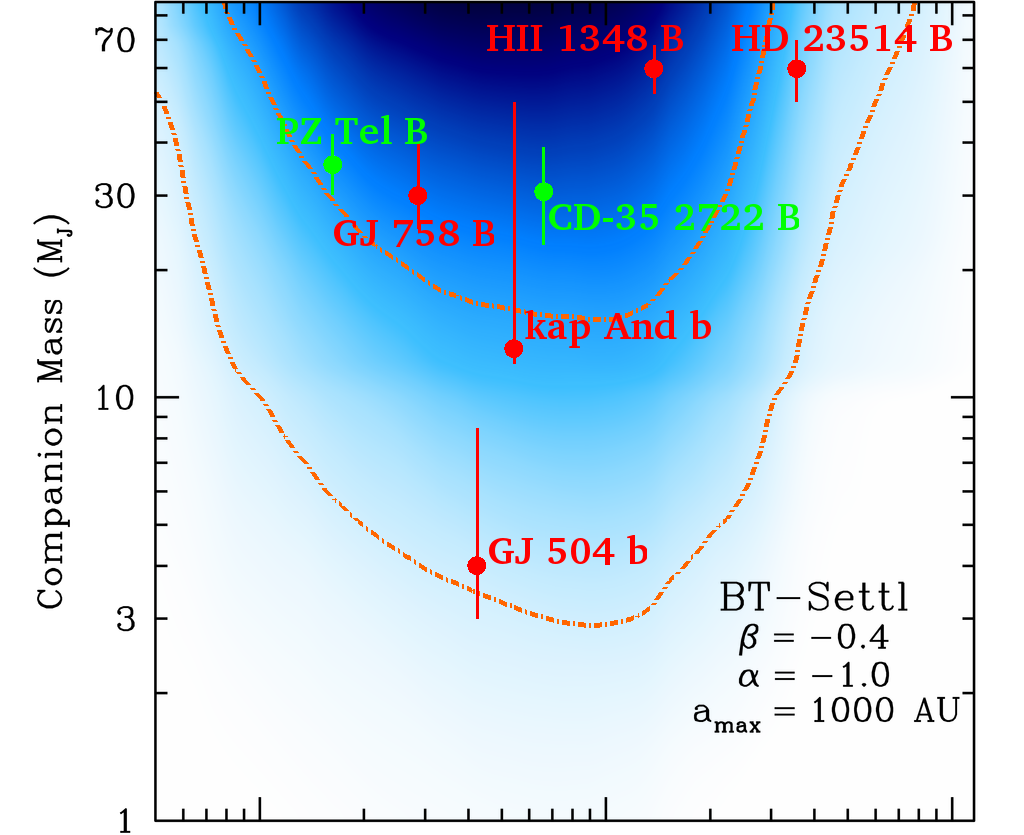}
\includegraphics[width=\linewidth]{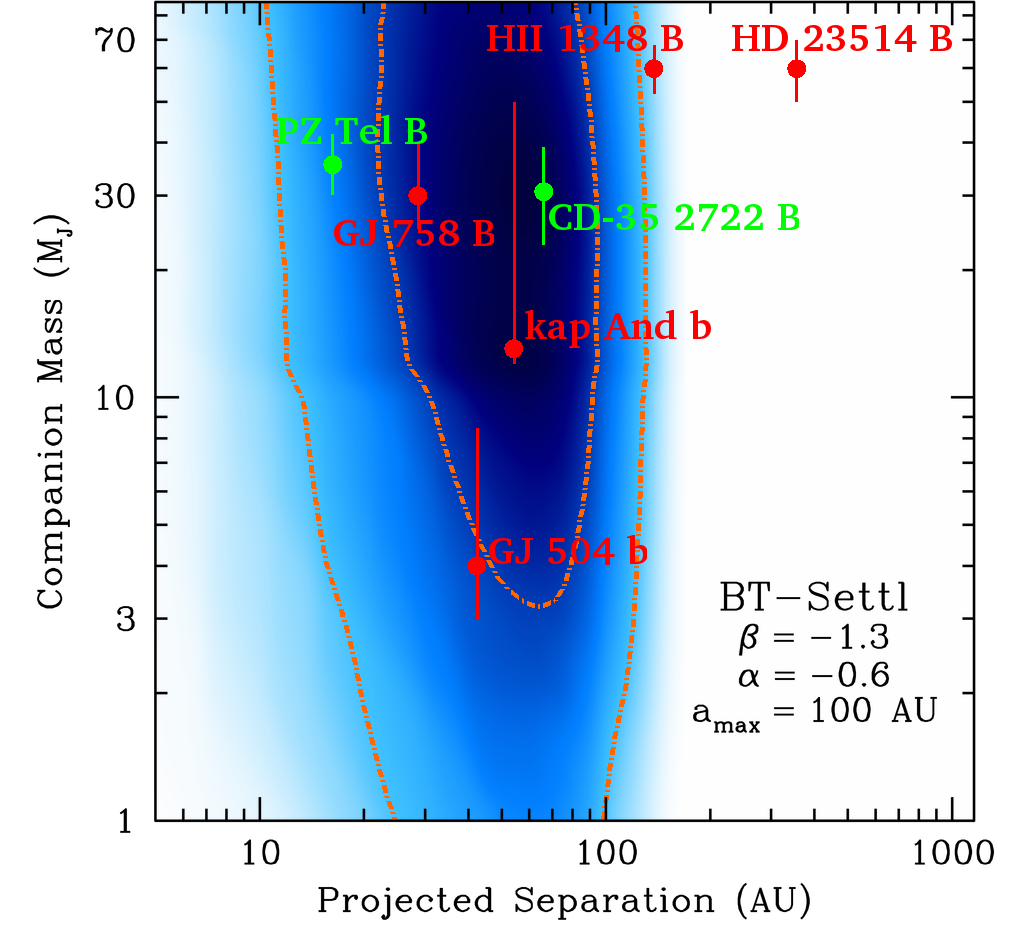}
\caption{Probability distributions, $dp/d\log M / d\log D$, under two different substellar distribution functions, $p(M, a) \propto a^{\alpha} M^{\beta}$.  NICI discoveries are shown in green; companions imaged by HiCIAO are shown in red.  The distribution with $\beta = -0.4$, \citep[top panel, taken from ][]{Metchev+Hillenbrand_2009}, predicts a large number of massive brown dwarfs at separations of $\sim$50--100 AU.  The bottom distribution is that published in \cite{Cumming+Butler+Marcy+etal_2008}, with $\alpha=-0.6$ and $\beta=-1.3$, extrapolated out to higher masses and semimajor axes, and assuming planet frequency to scale with host stellar mass.  The massive Pleiades brown dwarfs, {\sc H\,ii} 1348B and HD 23514B. would force the outer limit to extend to $\sim$300 AU, which is excluded by our sample (Figure \ref{fig:a_limits}).  They cannot arise from an extrapolation of the RV distribution function. } 
\label{fig:p_dists}
\end{figure}

We first show the predictions of two published distribution functions: $dN/dMda \propto M^{-0.4}a^{-1}$, derived for both stellar and substellar companions from $\sim$30--1500 AU \citep{Metchev+Hillenbrand_2009}, and $dN/dMda \propto M^{-1.3}a^{-0.6}$, derived for $\sim$1--10 $M_{\rm Jup}$ RV-detected companions from $\sim$0.03--3 AU \citep{Cumming+Butler+Marcy+etal_2008}.  In the former case, we extend the distribution down to 1 $M_{\rm Jup}$ and out to 1000 AU, well outside the field-of-view around nearly all of our targets.  In the latter case, we extrapolate the distribution function up to the hydrogen burning limit of $\sim$70 $M_{\rm Jup}$ and out to 100 AU, roughly the outermost semimajor axis consistent with our analysis in Section \ref{subsec:rv_limits}, and scale companion frequency with stellar mass.  
Figure \ref{fig:p_dists} shows the predicted probability densities, $dp/d\log M / d\log D$, for both of these distributions, together with contours of constant $dp/d\log M / d\log D$ enclosing 68\% and 95\% of the predicted detections.  The five HiCIAO detections, including GJ 504b and GJ 758B, are in red, while the two NICI detections are shown in green.

The \cite{Metchev+Hillenbrand_2009} distribution, the top panel in Figure \ref{fig:p_dists}, appears to provide a reasonably good fit to our sample, though it has some difficulty accounting for objects like GJ 504b (depending on its age).  The inclusion of intermediate separation low-mass companions from other surveys, like HD 95086b (which was discovered in a survey that did not detect any massive brown dwarfs), or of $\beta$ Pic b and HR 8799bcde, would add to this tension.  The RV-inspired distribution, bottom panel of Figure \ref{fig:p_dists}, terminates at 100 AU and is unable to account for the massive ($\sim$60 M$_{\rm J}$), long-period brown dwarfs discovered in the Pleiades.  We note that the distribution advocated by \cite{Zuckerman+Song_2009}, with $p(M, a) \propto M^{-1.2} a^{-1}$ (not shown), has a nearly identical mass dependence but extends to larger semimajor axes.  This distribution can accommodate all of the detections, though only if we extend it well below the $\sim$12 $M_{\rm Jup}$ limit suggested by \citeauthor{Zuckerman+Song_2009} in order to match GJ 504b.  

We now return to our sample, using our detections and contrast curves to constrain the substellar distribution function.  Given an assumed form of the distribution function, including the power law indices and mass and semimajor axis limits, we may then use the likelihood function (in the form of Equation \eqref{eq:log_like2}) to compare the distribution to our actual detections.  We assume uniform priors on the power law indices $\alpha$ and $\beta$, integrate the likelihood function, and treat it as a posterior probability distribution.  The additional effect of including GJ 504b and GJ 758B in a full analysis may be crudely estimated by scaling up our sample size to qualitatively account for the as-yet-unpublished non-detections.  Such a scaling would simply multiply $\langle N_{\rm obj} \rangle$ in Equation \eqref{eq:log_like2} by a constant, dropping out when constraining $\alpha$ and $\beta$.  This implicitly assumes that the ages, distances, and masses probed by the as-yet-unpublished non-detections are similar to those of the sample presented here.  Given the heterogeneity of both our combined sample and the unpublished SEEDS data, this is not a bad approximation.  We set $a_{\rm min} = 1$ AU and $a_{\rm max} = 1000$ AU, well outside the regions of interest, and use a lower mass cutoff of 5 M$_{\rm J}$, appropriate to a gravitational collapse scenario.

Including all of the detections in our merged sample, and adding GJ 504b and GJ 758B. we obtain a best-fit distribution function $p(M, a) \propto M^{-0.65 \pm 0.60} a^{-0.85 \pm 0.39}$ (1$\sigma$ errors).  Abandoning the lower limit on companion mass favors a distribution function with somewhat more high-mass objects, with the mass exponent becoming $-0.4 \pm 0.5$ (1$\sigma$ errors).  However, as discussed in Section \ref{sec:distributions}, there are theoretical reasons to impose such a lower mass limit for gravitational collapse, and a cutoff is also suggested by the dearth of companions $<$5 M$_{\rm J}$ in other high-contrast surveys.  

The normalization of the distribution function is given by a gamma distribution at fixed $\alpha$ and $\beta$; its maximum likelihood value produces five detections (the observed number) in our sample of 248 stars.  With $(\alpha, \beta) = (-0.85, -0.65)$, the maximum likelihood normalization constant gives $1.7$\% of stars with substellar companions between 5 and 70 $M_{\rm Jup}$ and between 10 and 100 AU.  The gamma distribution is asymmetric; the 68\% and 95\% confidence intervals are 1.2--2.8\% and 0.74--3.9\%, respectively, of stars with companions in the given mass and semimajor axis range.
If we also include uncertainty in $\alpha$ and $\beta$ by integrating the likelihood function and treating it as a posterior distribution, the uncertainty in the normalization constant increases somewhat.  For companions between 5 and 70 $M_{\rm Jup}$ and between 10 and 100 AU, the 68\% and 95\% confidence intervals become 1.0--3.1\% and 0.52--4.9\% of stars, respectively.

We also extrapolate our distribution out to $a=1600$ AU to facilitate comparison with \cite{Metchev+Hillenbrand_2009}, again using both the gamma distribution and uncertainties in $\alpha$ and $\beta$ to derive the full probability distribution of the fraction of stars hosting companions.  We find that, at 68\% confidence, 1.8--6.2\% of stars host brown dwarfs between 12 and 72 $M_{\rm Jup}$ and between 28 and 1600 AU (0.92--11\% at 95\% confidence).  These results agree very well with the \citeauthor{Metchev+Hillenbrand_2009} value of $3.2^{+3.1}_{-2.7}$\% (2$\sigma$ limits).  The latter analysis fixed $\alpha=-1$ and assumed a continuous mass function extending to stellar companions, making the agreement particularly gratifying.  Omitting the uncertainty in $\alpha$ and $\beta$, our results imply that 1.4--7.2\% of stars host 12--72 $M_{\rm Jup}$ with semimajor axes from 28--1600 AU at 95\% confidence.

\begin{figure}
\includegraphics[width=\linewidth]{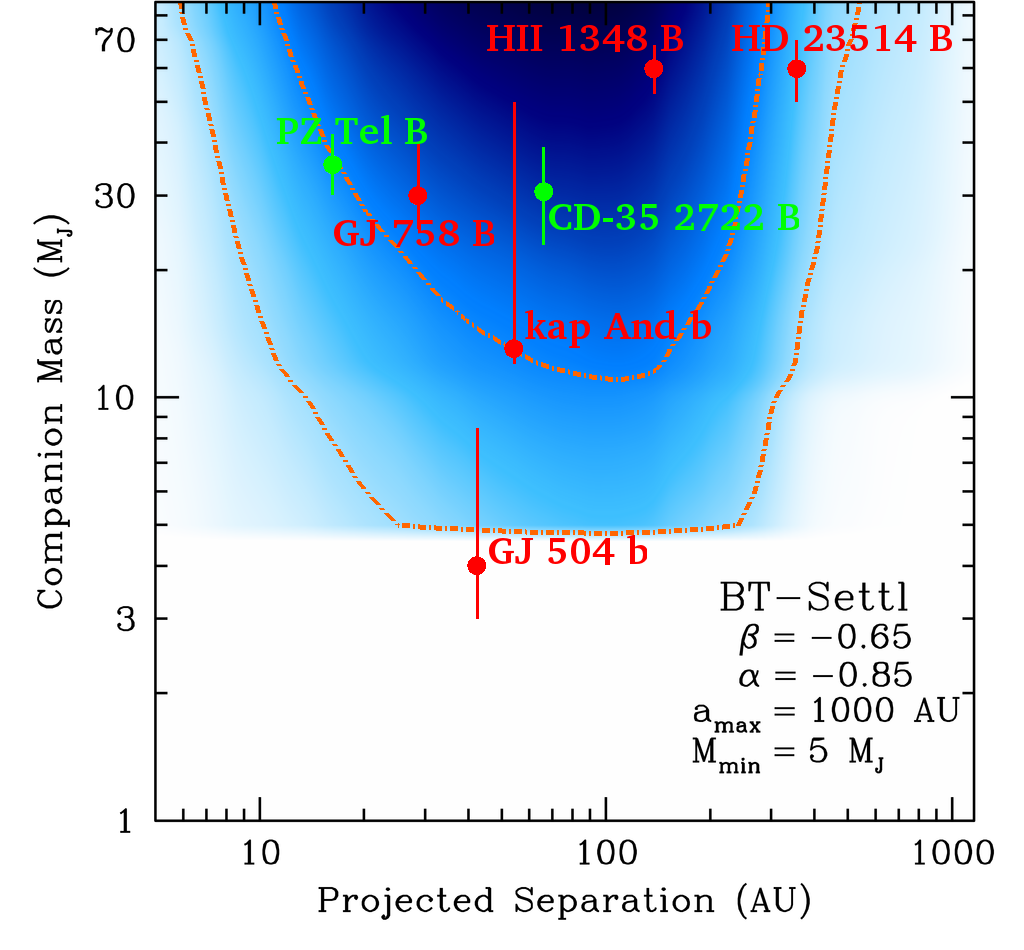}
\caption{The probability distribution, $dp/d\log M / d\log D$, for the maximum likelihood power law distribution including all seven detections shown: $p(M, a) \propto M^{-0.65 \pm 0.60} a^{-0.85 \pm 0.39}$ (1$\sigma$ errors).  NICI discoveries are in green, while companions imaged by HiCIAO are in red.  This result should be interpreted with caution, but is compatible with the distributions published in both \cite{Metchev+Hillenbrand_2009} and \cite{Zuckerman+Song_2009} within 1$\sigma$.  It suggests that we cannot reject a single substellar distribution function extending from massive brown dwarfs to massive exoplanets based solely on the sample presented here.  } 
\label{fig:p_dist_maxlike}
\end{figure}

The results from many other surveys in the $H$ and $K$ bands would fit nicely on Figure \ref{fig:p_dist_maxlike}, and further suggest a smooth distribution across the deuterium burning threshold.  \cite{Chauvin+Lagrange+Bonavita+etal_2010} reported the detection of three substellar objects in their survey, of which two, GSC-08047-00232B \citep{Chauvin+Lagrange+Lacombe+etal_2005}, and AB Pic b \citep{Chauvin+Lagrange+Zuckerman+etal_2005}, are $\sim$20 and $\sim$15 M$_{\rm J}$ brown dwarf companions 250--300 AU from their host stars.  
These data suggest that a smooth distribution extending all the way from massive brown dwarfs down to a theoretically motivated cutoff at $\sim$5 M$_{\rm J}$ is capable of explaining the vast majority of wide-separation companions below or near the deuterium-burning limit nominally separating planets from brown dwarfs.

\subsection{Limitations}

Our statistical analysis makes several assumptions that are not true in detail.  For example, we do not distinguish between single stars and binaries (which make up a relatively small fraction of our sample, $\sim$10\%).  Some orbits are unstable around binaries \citep{Holman+Wiegert_1999}; however, simply excluding these regions of parameter space and considering the rest other orbits to be as probable as around a single star is not a well-motivated solution.  Low-mass companions could be scattered to distant orbits by a close stellar binary, possibly making binaries better systems to find substellar objects.  In most cases, the orbital elements of the stellar binary are unknown anyway.  We therefore accept that our neglect of binarity might introduce a modest bias, but we lack a good solution other than excluding binaries altogether.

Our analysis also neglects sample biases, which can appear in many ways.  Stars hosting debris disks may be more likely to harbor planets.  Treating them identically to stars without infrared excesses, as we do here, could artificially depress the derived planet frequency.  SEEDS also avoided including known planet hosts and attempted to avoid observing stars targeted by other high-contrast instruments.  A tendency to avoid observing the same targets as other surveys, but to repeat some targets around which nothing was found, would likewise bias us against finding planets.

The choice of substellar cooling model introduces another opportunity for bias, as we discuss in Section \ref{sec:cooling_models}.  This is somewhat mitigated for our sample by the wide range of ages we probe, and by the fact that the luminosity of a brown dwarf or exoplanet depends so strongly on its mass and age.  Particularly for objects formed by direct gravitational collapse, there is little doubt about the initial thermodynamic state, though systematic errors could still arise from, e.g., a poor treatment of cloud formation.

Possibly the most serious limitation of our analysis comes from our ignorance of the formation mechanism of substellar objects, and our resulting assumption of separable power law distribution functions.  Substellar objects do not form in isolation; many have been imaged around stellar binaries.  These objects would interact dynamically with other stars or brown dwarfs, with lighter companions being preferentially scattered outwards, and more massive companions scattered inwards.  As a result, the initial and final companion mass distributions may be very different from one another.  Models of brown dwarf formation \citep[e.g.][]{Stamatellos+Whitworth_2009, Bate_2009} are beginning to predict distribution functions, but have difficulty producing low-mass brown dwarfs at moderate separations.  As a result, we have little choice but to proceed with simplistic models based only loosely on theoretical considerations.

In addition to these limitations, there are several objects not included in our sample that may be difficult to incorporate into a single substellar distribution like the one suggested in Section \ref{subsec:onedist}, and that may represent the high-mass, wide-separation tail of a core-accretion population.  These objects, including HR 8799bcde, $\beta$ Pic b, and HD 95086b. tend to occur around relatively massive A-type stars.  A stars rotate very rapidly and are therefore poor targets for radial velocity searches, making the application of a distribution function like that measured by \cite{Cumming+Butler+Marcy+etal_2008} (used in Section \ref{subsec:rv_limits}) an extrapolation in both separation and host stellar mass.  The masses of $\beta$ Pic b and at least some of the HR 8799 planets have been dynamically constrained to be $\lesssim$$10$ M$_{\rm J}$ \citep{Lagrange+DeBondt+Meunier+etal_2012, Dawson+Murray-Clay+Fabrycky_2011, Sudol+Haghighipour_2012}.  Such low-mass companions at $\sim$10--40 AU are difficult to form in-situ by direct gravitational collapse \citep{Rafikov_2005, Kratter+Murray-Clay+Youdin_2010}, while some recent studies have found that core-accretion may be viable out to a few tens of AU \citep{Lambrechts+Johansen_2012, Kenyon+Bromley_2009}.  Planet-planet scattering and migration do add some uncertainty to this picture.

As more low-mass companions are discovered, they may reveal a clear bimodal distribution in planet/star mass ratio, hints of which were shown by \cite{Currie+Burrows+Itoh+etal_2011}.  Such a clear separation would be strong evidence of different formation mechanisms.  Even stronger evidence would be planetary chemical compositions differing strongly from those of their host stars.  \cite{Konopacky+Barman+Macintosh+etal_2013} found signs of an enhanced C/O ratio in HR 8799c, though these results depend on chemical modeling of the atmosphere and were at modest significance.  A bimodal distribution of metal and/or carbon enhancements, with a strong correlation between composition, separation, and mass ratio, would probably be conclusive.  New and upcoming instruments like GPI \citep{Macintosh+Graham+Palmer+etal_2008}, SPHERE \citep{Beuzit+Feldt+Dohlen+etal_2008}, and CHARIS \citep{Peters+Groff+Kasdin+etal_2012} may be able to provide these data over the next few years.

\section{Conclusions} \label{sec:conclusions}

In this work, we present an analysis of high-contrast imaging of nearly 250 stars.  The targets span a wide range of spectral types and ages, and are composed of three published subsets of the SEEDS survey on the Subaru telescope, combined with the GDPS and the NICI MG sample.  We perform a uniform, Bayesian analysis of the ages of all of our targets, with determinations that are often more conservative than those adopted in previous papers.  Our sample includes five detected substellar companions.  Two of these are $\sim$60 M$_{\rm J}$ brown dwarfs around stars in the Pleiades; the others are a $\sim$13--50 M$_{\rm J}$ companion to the late B star $\kappa$ And, a $\sim$35 M$_{\rm J}$ companion to the late G star PZ Tel, and a $\sim$30 M$_{\rm J}$ companion to the early M star CD$-$35 2722.  

Our analysis includes a new method for calculating the likelihood function of a substellar distribution function by performing integrals analytically or evaluating them from tables whenever possible.  This represents a large improvement in efficiency over using Monte Carlo to evaluate completeness, and allows us to efficiently compute the likelihood of a wide range of models.  We use these techniques to compute the limits beyond which the distribution function measured for RV planets cannot be extended, finding a model-dependent maximum semimajor axis limit of $\sim$30--100 AU, similar to previous results.  However, we argue that the division of substellar objects at the deuterium burning limit is arbitrary, particularly in light of new discoveries that straddle that boundary, and we seek to model the entire substellar population using a single distribution function.  

Finally, we use Monte Carlo to compute the likelihood function of a unified substellar distribution function, including the five companions detected in our sample, plus an additional two objects, GJ 758B. and GJ 504b. discovered by HiCIAO, for a total of seven.  The inclusion of these objects does not bias the results as long as the distributions of target stars and contrast curves presented here are good matches to those of the unpublished non-detections from which GJ 758 and GJ 504 were culled.  Given the wide range of stellar properties and ages in both samples, this is a fairly good assumption.  With this caveat, we find that a single, separable power law, $p(M, a) \propto M^{-0.65 \pm 0.60} a^{-0.85 \pm 0.39}$ (1$\sigma$ errors), truncated at a theoretically motivated minimum mass of $\sim$5 M$_{\rm J}$, can account for the entire range of substellar companions detected in SEEDS.  The normalization of this distribution implies that, at 68\% confidence, 1.0--3.1\% of stars have substellar companions between 5 and 70 $M_{\rm Jup}$ and between 10 and 100 AU (0.52--4.9\% at 95\% confidence).  Extrapolating to larger separations, 1.8--6.2\% of stars (at 68\% confidence) have companions between 12 and 72 $M_{\rm Jup}$ and between 28 and 1600 AU (the limits used by \citealt{Metchev+Hillenbrand_2009}); or 0.92--11\% at 95\% confidence.  Our normalization is in excellent agreement with the \cite{Metchev+Hillenbrand_2009} result, that $3.2^{+3.1}_{-2.7}$\% (2$\sigma$ limits) of stars have brown dwarf companions within these limits.  

Our results suggest that many, perhaps most, of the substellar companions currently being discovered near and somewhat below the deuterium burning limit may share a common origin with more massive brown dwarfs.  Such objects would almost certainly form by gravitational collapse, either in a disk or in a fragmenting cloud.  There is currently little reason to consider the substellar companions, at least in our combined sample, to be the high-mass, long-period tail of the RV planet distribution.  

Upcoming surveys using instruments like GPI \citep{Macintosh+Graham+Palmer+etal_2008}, SPHERE \citep{Beuzit+Feldt+Dohlen+etal_2008}, and CHARIS \citep{Peters+Groff+Kasdin+etal_2012} will dramatically improve our sensitivity to low-mass companions, a region of parameter space that should be richly populated if current detections are described by an RV-like distribution function.  The discovery of many such objects, below the mass limits at which clouds and disks are expected to fragment, could point to an alternative formation scenario, like core-accretion followed by dynamical evolution.  If, however, such objects turn out to be exceptionally rare, the current population of directly imaged exoplanets likely represents the low-mass tail of the brown dwarfs.

\acknowledgments{This research is based on data collected at the Subaru Telescope, which is operated by the National Astronomical Observatories of Japan.  TDB gratefully acknowledges support from the Corning Glass Works Foundation through a fellowship at the Institute for Advanced Study.  This research has been supported in part by the World Premier International Research Center Initiative, MEXT, Japan.  MK acknowledges support from a JSPS Research Fellowship for Young Scientists (grant No.~25-8826).  JC was supported by the U.S. National Science Foundation under Award No.~1009203.  This research has made use of the SIMBAD database and Vizier service, operated at CDS, Strasbourg, France.  The authors wish to recognize and acknowledge the very significant cultural role and reverence that the summit of Mauna Kea has always had within the indigenous Hawaiian community.  We are most fortunate to have the opportunity to conduct observations from this mountain.}

\appendix

\section{Projected Separation Distribution Function}

Denoting the projected separation (in physical units) as $D$, and the ratio of the projected distance to the semimajor axis by $s \in [0,\,2]$, the probabillity distribution of $D$ is
\begin{equation}
p(D) = \int_{s_1}^{s_2} ds \, p(s) \, p \left( a = \frac{D}{s} \right) \frac{1}{s}~,
\end{equation}
where the last factor $1/s$ accounts for the volume element.  If $p(a)$ is a power law $a^\alpha$, truncated at $a_{\rm min}$ and $a_{\rm max}$, we have $s_1 = {\rm min}(2, D/a_{\rm max})$ and $s_2 = {\rm min}(2, D/a_{\rm min})$.  In the special case that the power law is truncated well outside the separations of interest, $s_1 \approx 0$ and $s_2 \approx 2$, and the dependence of $p(D)$ on the eccentricity distribution drops out altogether.

We can empirically derive $p(s)$ from $p(e)$ using the method described in the appendix of \cite{Brandeker+Jayawardhana+Khavari+etal_2006}.  This method assumes only that companions are observed at random times in their orbits and that their orbits are randomly oriented as seen from Earth.  \citeauthor{Brandeker+Jayawardhana+Khavari+etal_2006} suggest an eccentricity distribution $p(e) = 2e$, the theoretical distribution expected if the phase space density of companions is a function of energy only \citep{Ambartsumian_1937}.  However, planet-planet scattering disfavors eccentricities close to 1, producing distributions closer to the Rayleigh distribution with $\sigma \sim 0.3$ \citep{Juric+Tremaine_2008}.  Other authors \citep{Cumming+Butler+Marcy+etal_2008} have suggested a uniform distribution in eccentricity out to $e_{\rm max} \sim 0.8$.  We adopt the latter distribution, noting, however, that a Rayleigh distribution with $\sigma = 0.3$ produces nearly indistinguishable results.  

\begin{figure}
\centering\includegraphics[width=0.55\linewidth]{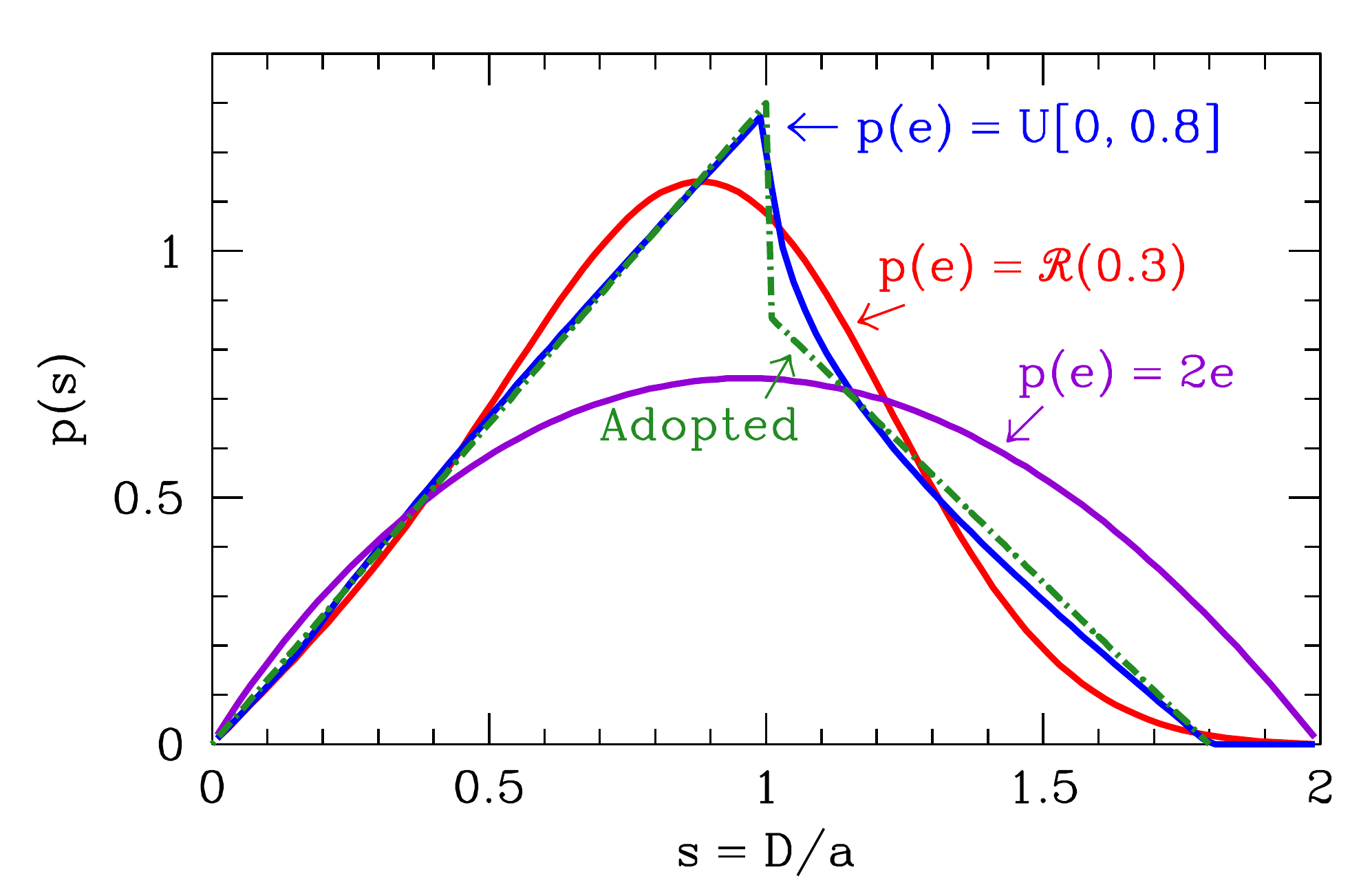} \\
\centering\includegraphics[width=0.55\linewidth]{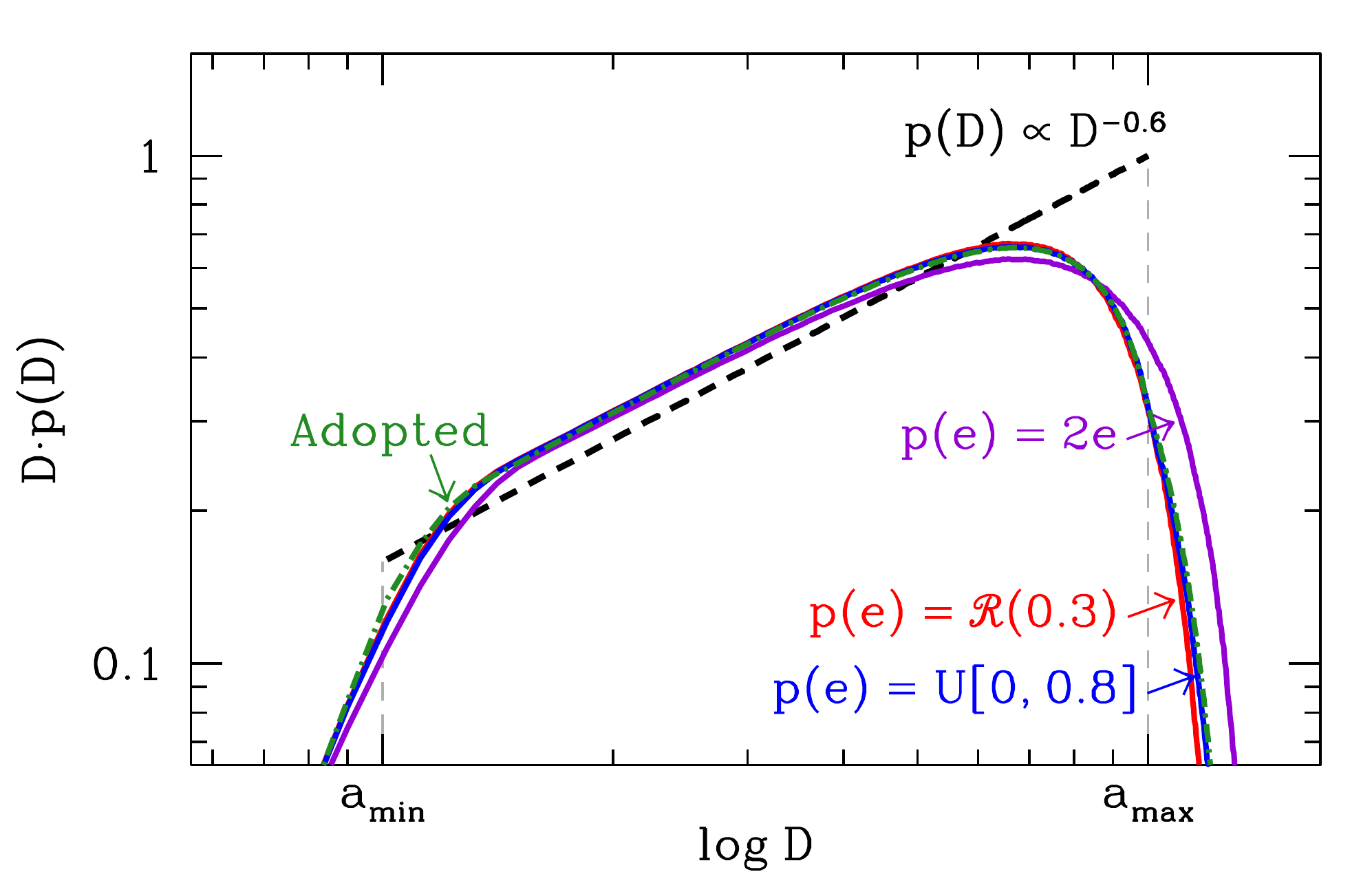}
\caption{Top panel: the distribution of $s = D/a$ assuming the eccentricity to be uniformly distributed between 0 and 0.8 (blue curve), Rayleigh distributed with $\sigma=0.3$ (red curve), or with $p(e)=2e$ (violet curve).  We adopt the piecewise linear approximation represented by the green curve.  Bottom panel: the distributions of projected separation produced by a truncated power law distribution in semimajor axis (black curve) for each distribution in $s$ from the top panel.  We use the piecewise analytic green dot-dashed curve, which is nearly identical to the results from the Rayleigh and uniform distributions; we also provide an expression that is nearly indistinguishable from the violet curve.}
\label{fig:e_dists}
\end{figure}

The top panel of Figure \ref{fig:e_dists} shows all three distributions in $s$.  The uniform distribution up to an eccentricity of 0.8 is well-fit by a piecewise linear function:
\begin{equation}
p(s) \approx 
\begin{cases}
1.3s & 0 \leq s \leq 1 \\
-\frac{35}{32} \left( s - \frac{9}{5} \right) & 1 < s < 1.8
\end{cases}
\end{equation}
Assuming $p(a) = a^{\alpha}$ for $a_{\rm min} \leq a \leq a_{\rm max}$, we then perform the integral 
\begin{equation}
p(D) = \int_{s_1}^{s_2} ds\,p(s)\,p\left(a = \frac{D}{s} \right) \frac{1}{s}~,
\end{equation}
with $s_1 = {\rm min}(D/a_{\rm max}, 1.8)$ and $s_2 = {\rm min}(D/a_{\rm min}, 1.8)$.  We have, after a bit of algebra,
\begin{equation}
p(D) = 
\begin{dcases}
D \left[ \frac{1.3}{1 - \alpha} \left( a_{\rm min}^{\alpha - 1} - a_{\rm max}^{\alpha - 1} \right) \right] & D < a_{\rm min} \\
D^{\alpha} \left[ \frac{315 + 68\alpha}{160\alpha (1 - \alpha)} \right] - D \left[ \frac{1.3}{1 - \alpha} a_{\rm max}^{\alpha - 1} + \frac{35}{32(1 - \alpha)} a_{\rm min}^{\alpha - 1} \right] - \frac{63}{32\alpha} a_{\rm min}^{\alpha} & a_{\rm min} \leq D < 1.8 a_{\rm min} \\
D^{\alpha} \left[ \frac{68\alpha + 315(1-1.8^{-\alpha})}{160 \alpha (1 - \alpha)} \right] - D \left[ \frac{1.3}{1 - \alpha} a_{\rm max}^{\alpha - 1} \right] & 1.8a_{\rm min} \leq D < a_{\rm max} \\
D^{\alpha} \left[ -\frac{35}{32} \frac{1.8^{1 - \alpha}}{\alpha(1 - \alpha)} \right] + 
    D \left[ \frac{35}{32} \frac{a_{\rm max}^{\alpha - 1}}{1 - \alpha} \right] +
    \frac{63}{32\alpha} a_{\rm max}^{\alpha} & a_{\rm max} \leq D \leq 1.8 a_{\rm max} 
\end{dcases}
\label{eq:D_dist_full}
\end{equation}
Given $a_{\rm min}$, $a_{\rm max}$, and $\alpha$, Equation \eqref{eq:D_dist_full} is a sum of piecewise power laws, and is trivial to integrate analytically.  The bottom panel of Figure \ref{fig:e_dists} shows the final distributions $p(D)$ for each of the eccentricity distributions we consider, together with the piecewise analytic approximation given by Equation \eqref{eq:D_dist_full}.  The blue and green curves, representing the exact result and our approximation, are nearly indistinguishable from one another and from the Rayleigh distribution (red curve).

For completeness, we also work out the case of an eccentricity distribution with 
\begin{equation}
p(e) = 2e~,
\end{equation}
the result obtained assuming the phase space density to be a function only of orbital energy.  In this case, we can very closely approximate $p(s)$ by 
\begin{equation}
p(s) \approx \frac{3}{4} \left( 2s - s^2 \right)~.
\end{equation}
In their paper, \cite{Brandeker+Jayawardhana+Khavari+etal_2006} used a sine curve, which provides a slightly worse fit and is much more difficult to handle analytically.  With our approximation, the distribution $p(D)$ becomes
\begin{equation}
p(D) = D^\alpha \left[ \frac{3}{2(1-\alpha)} \left(s_2^{1-\alpha} - s_1^{1-\alpha} \right)
    - \frac{3}{4(2-\alpha)} \left(s_2^{2-\alpha} - s_1^{2-\alpha} \right) \right]~,
\label{eq:D_dist_full_e2}
\end{equation}
with $s_1 = {\rm min}(D/a_{\rm max}, 2)$ and $s_2 = {\rm min}(D/a_{\rm min}, 2)$.

\section{Detections at a Given Separation}

Suppose that the noise at angular separation $D$ is $\sigma_D$, and the formal detection threshold is $L_{\rm lim} = N_\sigma \sigma_D$.  The probability of detecting a companion of luminosity $L$, assuming Gaussian errors, is then
\begin{equation}
p({\rm detect}|L) = \frac{1}{2} + \frac{1}{2} {\rm erf} \left(\frac{N_\sigma(L-L_{\rm lim})}{L_{\rm lim} \sqrt{2}} \right)~,
\label{eq:erf2}
\end{equation}
where erf is the error function.  The total number of detected companions at angular separation $D$ is then
\begin{equation}
N(D) = \int_0^\infty \frac{dN}{dL} p({\rm detect}|L)\,dL
= \int_0^\infty \frac{dN}{dM} \frac{dM}{dL} p({\rm detect}|L)\,dL~.
\end{equation}
Substellar cooling models are generally presented as grids in mass-luminosity space, which we interpolate using piecewise power laws, with 
\begin{equation}
\frac{L}{L_i} = \left( \frac{M}{M_i} \right)^{\gamma_i}
\end{equation}
for $M_{i,0} \leq M < M_{i, 1}$ or, equivalently, for $L_{i,0} \leq L < L_{i,1}$.  Defining 
\begin{equation}
x \equiv \frac{L}{L_{\rm lim}} ~~ {\rm and}~~ \Gamma_i \equiv (\beta + 1 - \gamma_i)/\gamma_i~,
\end{equation}
and assuming the distribution function of mass to be
\begin{equation}
p_M(M) \propto M^\beta~,
\end{equation}
we have, after some algebra,
\begin{align}
N(D) &= \sum_i M_i p_M (M_i) \int_{x_{i,0}}^{x_{i,1}} \frac{1}{\gamma_i} x^{\Gamma_i} \left[ \frac{1}{2} + \frac{1}{2} {\rm erf} \left( \frac{(x - 1)N_\sigma}{\sqrt{2}} \right) \right] dx~. 
\label{eq:int_n_D}
\end{align}
The integral in Equation \eqref{eq:int_n_D} is a function only of the limits of integration and of the quantity $\Gamma_i = (\beta + 1 - \gamma_i)/\gamma_i$.  We approximate the integral in different regimes depending on the values of $x_{i,0}$, $x_{i,1}$, and $\gamma_i$.  For $5 \leq N_\sigma \leq 6$, $x = 0.4$ corresponds to 3--$4 \sigma$ below the formal detection limit, while $x = 1.6$ is 3--$4\sigma$ above the formal detection limit.  We adopt the following approximations:
\begin{enumerate}
\item $x_{i,1} < 0.4$: The integral is nearly zero.
\item $x_{i,1} > 0.4$ and $\Gamma \notin [-6,\,10]$: For $\beta \in [-3,\,2]$ (ranging from exceptionally bottom heavy to exceptionally top-heavy; $\beta \sim -1$ from RV studies), this would imply $\gamma < 1/3$, i.e., the dependence of luminosity of mass is exceptionally weak.  We approximate the error function as a constant, obtaining
\begin{equation}
\frac{1}{\beta + 1} \left( M_{i,1} p_M (M_{i,1}) - M_{i,0} p_M (M_{i,0} \right) \left[ \frac{1}{2} + \frac{1}{2} {\rm erf}\left(\frac{\left(L_i/L_{\rm lim} - 1 \right) N_\sigma}{\sqrt{2}} \right) \right] ~.
\end{equation}
\item $0.4 < x_{i,1} < 1.6$ and $-6 \leq \Gamma \leq 10$: We evaluate the integral from tabulated quadratic fits.  We set the lower limit of integration to ${\rm min}(0.4,\,x_{i,0})$.
\item $x_{i,1} > 1.6$ and $-6 \leq \Gamma \leq 10$: We evaluate the integral from tabulated quadratic fits, setting the lower limit of integration to ${\rm min}(0.4,\,x_{i,0})$ and the upper limit to 1.6.  We then integrate the entire rest of the mass distribution (assuming a nearly monotonic mass-luminosity relation), obtaining
\begin{equation}
\frac{1}{\beta + 1} \left[ M_{\rm max} p_M (M_{\rm max}) - M_i p_M (M_i) \left( \frac{1.6 L_{\rm lim}}{L_i} \right)^{(\beta + 1)/\gamma_i} \right]~.
\end{equation}
\end{enumerate}
Assuming we pre-compute $p_M$ at the tabulated masses, each of these approximations requires at most one call to a special function (power or erf), a handful of array lookups, and $\sim$20 floating point operations.  Our tabulated fits to the integrals are always accurate to better than 0.1\% of the integral evaluated between $x=0.4$ and 1.6; these approximations therefore introduce less error than the (inevitable) interpolations over the grid of models.

\bibliographystyle{apj_eprint}
\bibliography{seeds_refs}

\end{document}